\documentclass[pre,floatfix,superscriptaddress]{revtex4}
\usepackage{amssymb,amsfonts,amsmath}
\usepackage{bbm,graphicx, mathrsfs, float, mathtools, cases,url}
\usepackage{hyperref}

\newcommand{\1}{\mathbbm 1}
\newcommand{\sign}{\mathrm{sign}}
\newcommand{\bx}{\mathbf x}
\newcommand{\bt}{\mathbf t}
\newcommand{\bu}{\mathbf u}
\newcommand{\bh}{\mathbf h}
\newcommand{\by}{\mathbf y}

\newcommand{\dd}{\partial}
\newcommand{\sm}{\setminus}

\begin{document}
\title{Large deviations of cascade processes on graphs}

\author{F.~Altarelli}
\affiliation{Department of Applied Science and Technology, Politecnico di Torino, Corso Duca degli Abruzzi 24, 10129 Torino, Italy}
\affiliation{Collegio Carlo Alberto, Via Real Collegio 30, 10024 Moncalieri, Italy}
\author{A.~Braunstein}
\affiliation{Department of Applied Science and Technology, Politecnico di Torino, Corso Duca degli Abruzzi 24, 10129 Torino, Italy}
\affiliation{Human Genetics Foundation, Via Nizza 52, 10126 Torino, Italy}
\affiliation{Collegio Carlo Alberto, Via Real Collegio 30, 10024 Moncalieri, Italy}
\author{L.~Dall'Asta}
\affiliation{Department of Applied Science and Technology, Politecnico di Torino, Corso Duca degli Abruzzi 24, 10129 Torino, Italy}
\affiliation{Collegio Carlo Alberto, Via Real Collegio 30, 10024 Moncalieri, Italy}
\author{R.~Zecchina}
\affiliation{Department of Applied Science and Technology, Politecnico di Torino, Corso Duca degli Abruzzi 24, 10129 Torino, Italy}
\affiliation{Human Genetics Foundation, Via Nizza 52, 10126 Torino, Italy}
\affiliation{Collegio Carlo Alberto, Via Real Collegio 30, 10024 Moncalieri, Italy}

\begin{abstract}
Simple models of irreversible dynamical processes such as Bootstrap Percolation have been successfully applied to describe cascade processes in a large variety of different contexts.
However, the problem of analyzing non-typical trajectories, which can be crucial for the understanding of the out-of-equilibrium  phenomena, is still considered to be intractable in most cases.
Here we introduce an efficient method to find and analyze optimized trajectories of cascade processes. We show that for a wide class of irreversible dynamical rules, this  problem can be solved efficiently on large--scale systems.
\end{abstract}


\maketitle

\section{Introduction}\label{sec1}
Large-scale cascading processes observed in physical and biological systems can be described and understood by means of stylized models of propagation on lattices or graphs.
Over the last forty years, these models have found application to problems arising in a number of different contexts, ranging from  competing interactions in dilute magnetic systems \cite{CLR79,DSS97}, jamming transitions in glass formers and granular media \cite{jamming}, epidemic spreading \cite{NE02}, activation cascades in cortical \cite{brain} and other biological networks \cite{bio} to the spread of information and innovations in social models \cite{S73,G78,W02,JY05,AOY11} and propagation of liquidity shocks in financial interbank lending networks \cite{EN01,HM11}.
In all these problems the basic units composing the systems are discrete and undergo irreversible transitions from an ``inactive'' state to an ``active'' one depending on the state of their neighbors. Following recent works in the computer science community \cite{K07}, we refer to this class of dynamical processes as models of {\em progressive dynamics}.

Theoretical works across several disciplines have focused mostly on the mechanisms responsible for the emergence of some collective behavior, explaining under which conditions, on the dynamical rule and the graph/lattice structure, large-scale propagations can be observed as an outcome of typical realizations of the process, i.e. when starting from random initial conditions. Because of the intrinsic non-linearity of the dynamics, a {\em critical (or tipping) point} usually separates a region of parameters in which the dynamics typically occurs only locally from a region of large-scale propagations. This is exactly what occurs in celebrated models of statistical physics, such as bootstrap and $k$-core percolation \cite{CLR79,DGM06,BP07,FSD07} and zero-temperature Ising-like models \cite{DSS97,OS10}, whose critical properties have been extensively studied for several classes of networks, such as $d$-dimensional lattices and random graphs. Similarly, tipping points are observed in simple models of binary decisions with externalities \cite{S73,W02,JY05}, providing an explanation for the occurrence of abrupt changes in the collective behavior of socio-economic systems. These analyses are usually performed either by simulating the evolution of the dynamical rule and averaging over many (randomly drawn) initial conditions, or by resorting to approximate descriptions of the dynamics in the form of differential equations based on mean-field and pair-approximation techniques \cite{BBV08,G11}.

While the average dynamical properties of these models starting from random initial conditions are rather well understood on general networks, their {\em large deviations}, describing macroscopic behaviors that deviate considerably from the average ones, is still a largely unexplored domain of research that goes beyond the means of current methods of analysis. Large deviations are of interest for at least two different reasons: because they correspond to {\em desired} final states (e.g. extraordinarily large propagations of a small set of initially active nodes) or because they correspond to an {\em observed} final state of an unknown initial one. The application of large deviation analyses to the non-equilibrium dynamics of interacting particle systems is subject of intense study in statistical physics \cite{T09}. Models of progressive dynamics offer a sufficiently simple, though non trivial, setting to extend these studies to systems with complex interaction patterns such as random graphs and complex networks.

In this paper we consider the problem of characterizing dynamical trajectories with interesting non-typical statistical properties in deterministic progressive models. In this class of models, the choice of the initial conditions completely determines the dynamical trajectory of the system. However, because of the non-linearity of the local update rule, even slight differences in the initial conditions can result in completely different collective behaviors. By averaging over all possible initial conditions or drawing them at random, the macroscopic quantities of interest are dominated by their typical behavior that can be extremely different from the observed one when a particular choice of the initial conditions is made. On the contrary, we will provide here a method to estimate the statistical properties of rare, but relevant, dynamical trajectories and find the initial conditions that give rise to cascading processes with some desired properties. Understanding under which conditions a rare large-scale propagation may occur and estimating the probability and other statistical properties of such an outcome have remarkable practical applications in a variety of fields beyond physics, such as the study of the spread of information in social networks, the problem of targeted silencing in gene regulatory networks, or the development of systemic risk measures and control techniques in financial and infrastructure systems.

Our approach is based on a static representation of the dynamical rules of deterministic progressive models that allows one to recast the study of their large deviations into the evaluation of a partition function. In networked systems this can be done by means of the cavity method and derived message-passing algorithms. Even though the method we propose is very general and can be applied to any deterministic progressive dynamics with discrete degrees of freedom, in the following we shall consider explicitly the Linear Threshold Model (LTM), a prototypical model for the analysis of cascade processes on networks.

\section{Typical behavior of progressive dynamics on graphs}\label{sec2}
In this section we will present the cavity formalism for the analysis of the typical behaviour of trajectories in progressive dynamics, from which we will recover previously known results; the formal connection with the main result on large deviations will be discussed in Appendix~\ref{BPothasasa}.
We consider a generic deterministic progressive dynamics in discrete time defined over a graph $G = (V, E)$ and involving discrete state variables $\mathbf x = \{x_i, i \in V\}$. For simplicity we shall assume that there are only two states, $x_i = 0$ called \emph{inactive} and $x_i = 1$ called \emph{active}, the generalization to more states being straightforward. A vertex which is active at time $t$ will remain active at all subsequent times, while a vertex which is inactive at time $t$ can get activated at time $t+1$ if some condition, depending on the state of its neighbors in $G$ at time $t$ and expressing the dynamical rule considered, is satisfied. For instance, in the Linear Threshold Model \cite{G78,AOY11,KKT03}, the dynamics is defined by the rule
\begin{equation}
 x_i^{t+1} =
 \begin{cases}
 1 & \text{if } x_i^t = 1 \text { or } \sum_{j \in \dd i} w_{ji} x_j^t \geq \theta_i \,, \\
 0 & \text{otherwise} \,
 \end{cases}
\end{equation}
where $w_{ij} \in \mathbbm R^+$ are weights associated to directed edges $(i,j) \in E$,  $\theta_i \in \mathbbm R^+$ are thresholds associated to $i \in V$ and $\dd i$ denotes the set of neighbors of $i$ in $G$. The model is strictly related to the zero-temperature limit of the random-field Ising model \cite{DSS97,OS10} and to the Bootstrap Percolation process \cite{CLR79,DGM06,BP07}. The active nodes at time $t=0$ are called the {\em seeds} of the progressive dynamics.

\begin{figure}
\includegraphics[width=0.6\columnwidth]{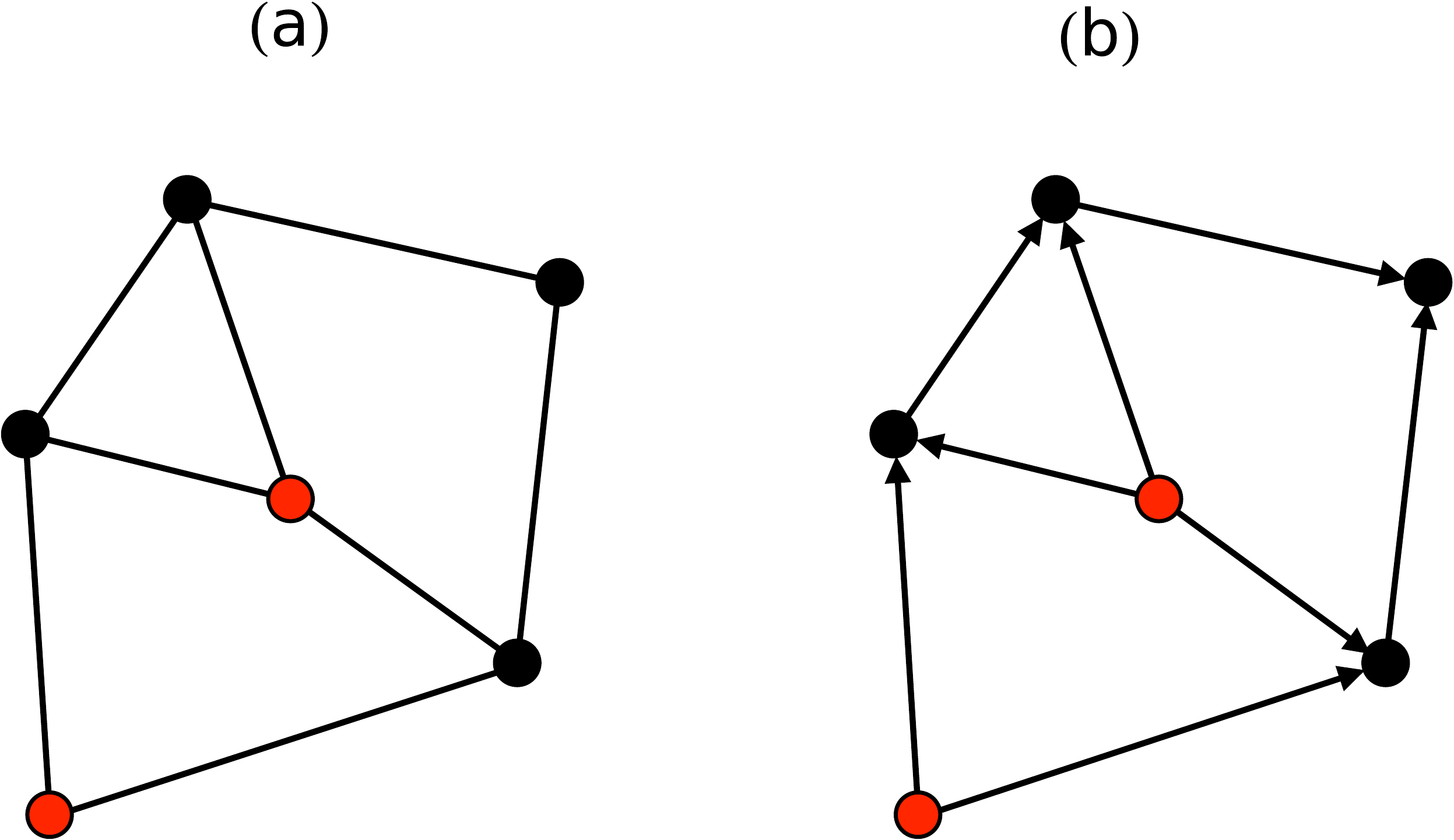}
\caption{(Color online) An example of the relation between the progressive models and directed acyclic graphs (DAG). A graph of $6$ vertices undergoes a LTM with two seeds (vertices marked in red). The weights on all edges are equal to $1$ and the threshold is equal to $2$ for every node. The result of the dynamics is the DAG on the right. The direction of the edges in the DAG represent the causal relations behind node activations.
\label{dag}}
\end{figure}

\subsection{The direct dynamical problem}\label{sec2a}
A peculiar property of a progressive process defined on a graph $G$ is that any realization of the process is in one-to-one correspondence with a directed acyclic subgraph of $G$. Let us consider a set of seeds and, for each time step $t$, draw a directed edge connecting the nodes activated at time $t$ to their neighbors activating at later times. The final result is a directed acyclic graph (DAG) as shown in Fig.\ref{dag} for an illustrative case. When the initial conditions are drawn from a distribution, the probability that a node $i$ is active is given by the probability that the node is in the set of seeds plus the probability that it is not a seed but it gets activated during the dynamics. The latter is the probability that $i$ is reached by directed paths from the seeds in the ensemble of DAGs associated to the initial distribution. When the underlying graph is a tree and the initial conditions are drawn from a product measure, i.e. with probability $Pr\{\bx^0=\bx\} = \prod_i p_i^{x_i} (1-p_i)^{1-x_i}$,  the probability $\rho_i(t)$ that a node $i$ is active at time $t$ can be computed exactly by a simple recursive approach. For instance, in the LTM, it is given by
\begin{eqnarray}\label{eq-rho}
 \rho_i^t & = & p_i + (1-p_i)\left\langle Pr\{ x_i^t=1 | x_i^0 = 0\} \right\rangle \\
\nonumber & = & p_i + (1-p_i)\sum_{\substack{I\subseteq \partial i \\ \sum_{\ell \in I}w_{\ell i} \geq \theta_i}}\prod_{\ell\in I} \chi_{\ell i}^t \prod_{k\in \partial i\setminus{I}}(1-\chi_{k i}^t) ,
\end{eqnarray}
with
\begin{equation}\label{eq-chi}
\chi_{j i}^{t+1} =  p_{j} + (1-p_j)\sum_{\substack{I\subseteq \partial j\setminus{i} \\ \sum_{\ell \in I}w_{\ell  j} \geq \theta_j}}  \prod_{\ell\in I}\chi_{\ell j}^t\prod_{k\in \partial j\setminus(\{i\} \cup I)}(1-\chi_{k j}^t).
\end{equation}
The quantity $\chi_{ji}^t$ is a cavity marginal expressing the probability that node $j$ is active at time $t$ in the absence of node $i$. Due to the nature of the process, assuming that node $i$ is absent is equivalent to assume that it is inactive, therefore the causal structure implied by Eqs.\eqref{eq-rho}-\eqref{eq-chi} is exact on the tree.
When the underlying graph has loops, the recursive equation is not exact. In this case, the DAG corresponding to a single dynamical evolution can present multiple directed paths connecting node $i$ to the same seed (see for instance Fig.\ref{dag}). In this case the decorrelation assumption behind \eqref{eq-rho}-\eqref{eq-chi} is not correct because two paths reaching $i$ from different neighbors could originate in the same seed and therefore they might be correlated. This argument shows that the activation probability of a node estimated by \eqref{eq-rho}-\eqref{eq-chi} is always an upper bound of the real one. Despite this limitations, the local tree-like approximation gives approximately correct results on sufficiently sparse graphs.

This cavity-like approach to study the time dependent dynamics of progressive processes on graphs was recently put forward, in slightly different contexts, by several authors. In particular, Ohta and Sasa have used a very similar approach to study the zero-temperature dynamics of the random-field Ising model on the Bethe lattice \cite{OS10}, while Karrer and Newman \cite{KN10} and No\"el et al. \cite{NAHDMD12} developed a similar method for the susceptible-infected model of epidemic spreading.

\begin{figure}
\includegraphics[width=0.6\columnwidth]{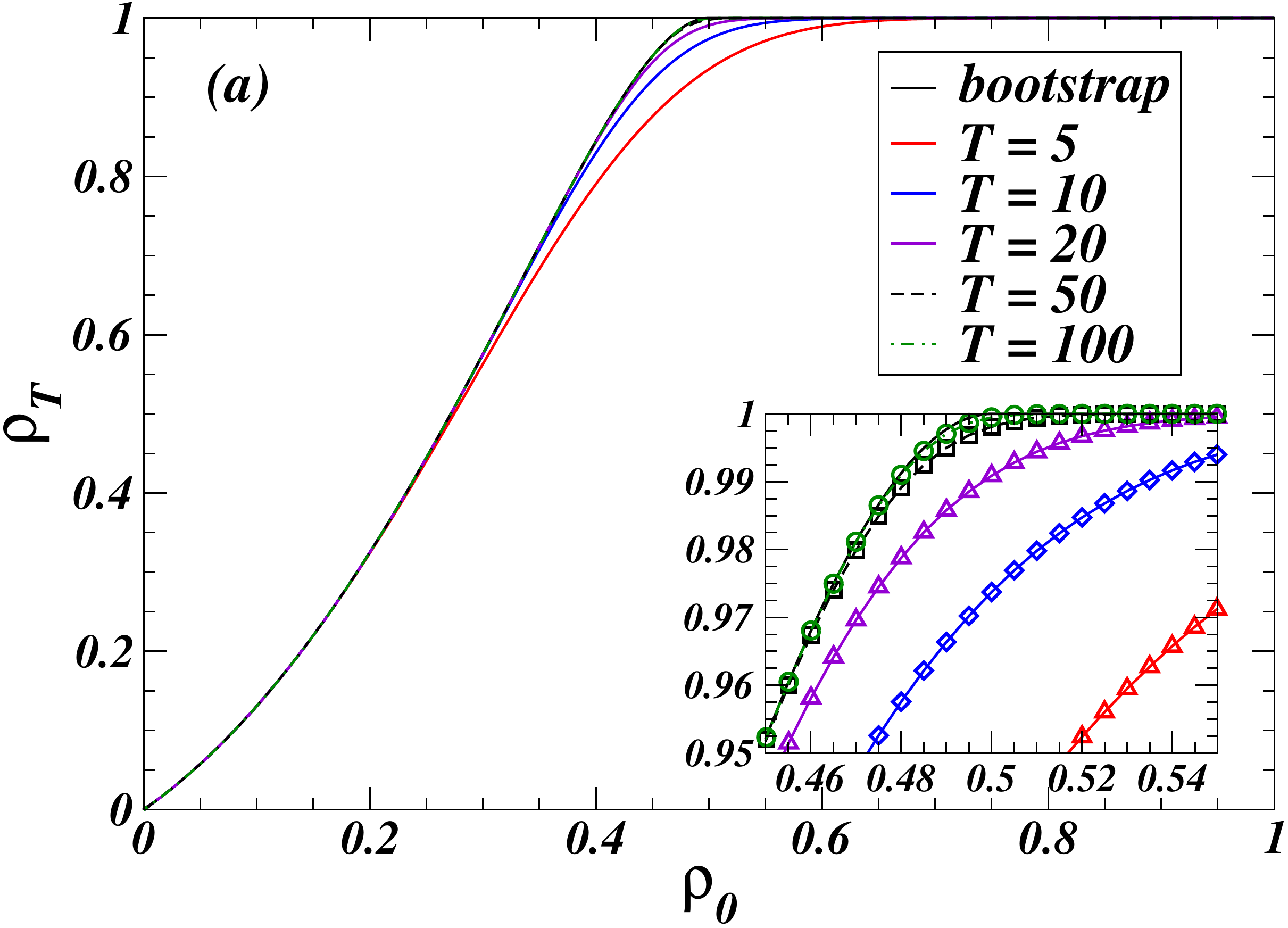}
\includegraphics[width=0.6\columnwidth]{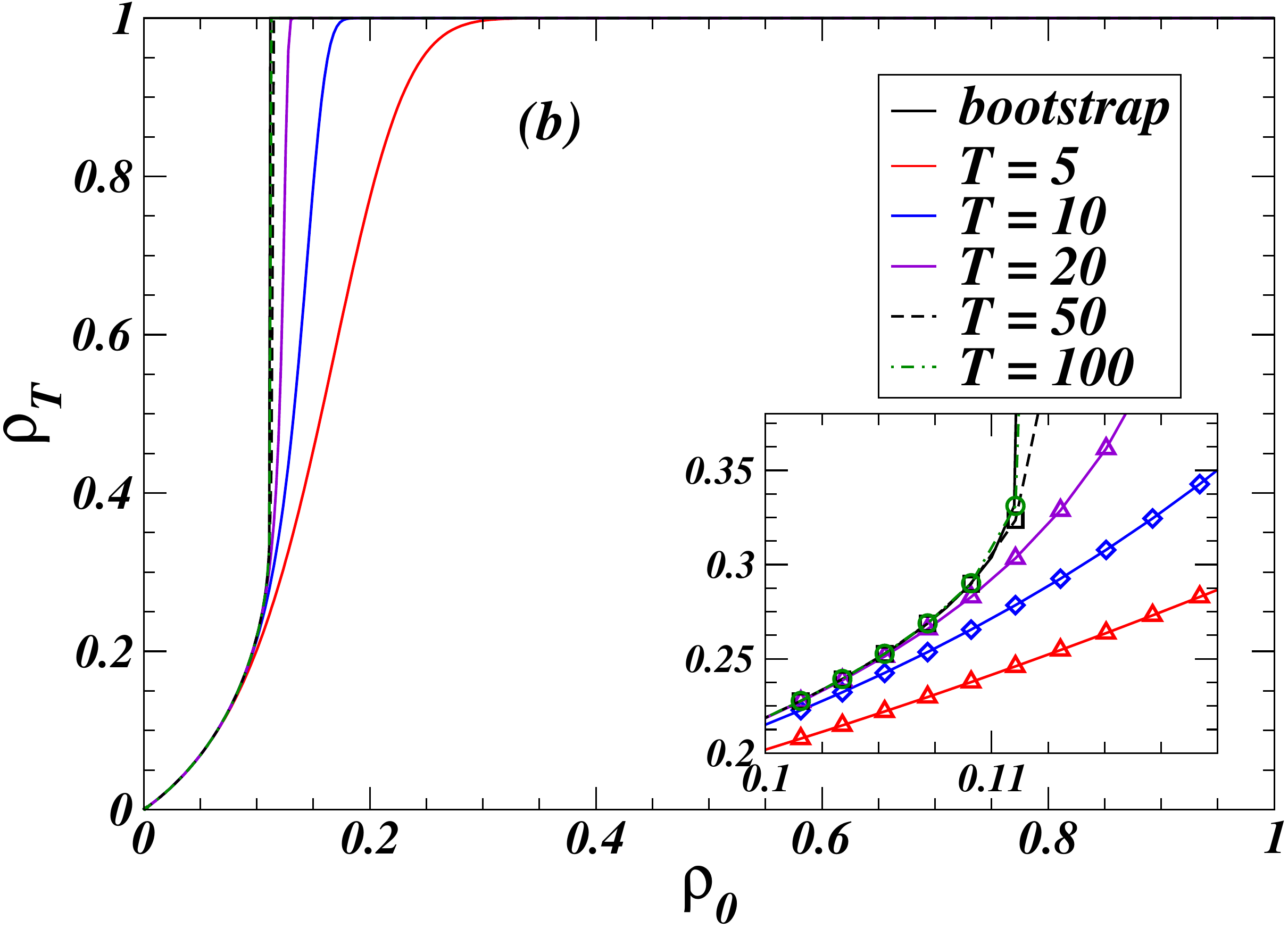}
\caption{(Color online) Plot $\rho_T$ v.s.$\rho_0$ obtained solving the time-dependent equations \eqref{eq-rho}-\eqref{eq-chi} and the Bootstrap Percolation equations \eqref{bootstrap1}-\eqref{bootstrap2} on regular random graphs of degree $K=3$ (a) and $K=4$ (b), for threshold $\theta = 2$. Increasing the duration $T=5,10,20,50, 100$ of the dynamics, the curves obtained using \eqref{bootstrap1}-\eqref{bootstrap2} get closer to the solution (from right to left) of the corresponding bootstrap percolation equations. The insets highlight the regions close to the activation transitions.
\label{figEps0VarT}}
\end{figure}

\subsection{Relation to the Bootstrap Percolation Problem}\label{sec2b}
In the Bootstrap Percolation \cite{CLR79}, the sites of an empty lattice are first randomly occupied with probability $q$, and then all occupied sites with less than a given number $m$ of occupied neighbors are successively removed until a stable configuration is reached. Like in standard percolation, in the limit of infinitely large graphs, the average properties of the model are characterized by the existence of a critical density $q_c$ of initially occupied sites below which the stable configuration of the system is the empty one. Taking $p= 1-q$ and interpreting empty (occupied) sites as active (inactive) nodes, the bootstrap percolation process can be mapped exactly on a LTM with uniform weights $w_{ij}=1, \forall (i,j)\in E$ and thresholds $\theta_i = k_i - m, \forall i \in V$, where $k_i$ is the degree of vertex $i$. We consider the simple case of a regular random graph with degree $K$ and uniform thresholds equal to $\theta$ for all nodes. Because of the homogeneity of the graph, one can assume all marginals $\chi_{ij}^t$ to be identical, obtaining for $\chi = \lim_{t \to \infty}\chi^t$ the self-consistent equation
\begin{equation}\label{bootstrap1}
\chi = p + (1-p) \sum_{n=\lceil\theta\rceil}^{K-1}\binom{K-1}{n} {\chi}^n {(1-\chi)}^{K-1-n}.
\end{equation}
The final density of active nodes when each node can be a seed with the same probability $p$ is given by
\begin{equation}\label{bootstrap2}
\rho = p + (1-p) \sum_{n=\lceil\theta\rceil}^{K}\binom{K}{n} \chi^n (1-\chi)^{K-n}.
\end{equation}
As expected, Eqs. \eqref{bootstrap1}-\eqref{bootstrap2} can be mapped exactly on the standard self-consistent equations derived for the bootstrap and $k$-core percolation processes on regular random graphs \cite{CLR79,DGM06,BP07}.
Figure \ref{figEps0VarT} displays the behavior of the final density $\rho_T$ of active nodes as function of the density $\rho_0$ of seeds in two interesting cases with $\theta=2$ and $K=3$ (a) and $K=4$ (b).  If we stop the dynamics at a finite number of time steps $T$, e.g.  $T=5$, the activation process is smooth in both cases. Increasing $T$, the curves for $K=4$ becomes much steeper than for $K=3$. In the limit $T\to \infty$, corresponding to the static bootstrap-like calculation, the two cases have completely different behavior: the activation transition is continuous for $K=3$ and discontinuous for $K=4$ (see the insets of Fig.\ref{figEps0VarT}). This is exactly the well-known critical phenomena observed in bootstrap percolation models on Bethe lattices \cite{CLR79,BP07}. In general, for $\theta=1, K-1$ the whole graph activates ($\rho_{\infty} = 1$) continuously at a finite density $\rho_{0}^c$, whereas for $1 < \theta < K-1$ the activation is abrupt. The generalization of the calculations to other values of the weights and thresholds as well as to non-regular uncorrelated random graphs is straightforward.

\section{Large Deviations of progressive dynamics on graphs}
In this Section we consider the inverse problem of dynamical evolution, i.e. the problem of finding the initial conditions that give rise to a desired final state. If we focus on the behavior of some macroscopic observable, such as the number of activated nodes in the final state as function of the number of seeds, the inverse problem corresponds to investigate the large deviation properties of the dynamics.

\subsection{The inverse dynamical problem}

Because of irreversibility, the trajectory $\underline{\bx}^T = \{\bx^0, \dots, \bx^T\}$ representing the time evolution of the system can be fully parametrized by a configuration $\bt = \{t_1, \dots, t_N\}$, where $t_i \in \mathcal T = \{0, 1, 2, \dots, T, \infty \}$ is the activation time of node $i$. We conventionally set $t_i = \infty$ if $i$ does not activate within an arbitrarily defined stopping time $T$. In general, if the number of possible single-node trajectories is $n$, we can use a discrete variable taking $n$ states.
Given a set of seeds $S=\{i: t_i=0\}$, the solution of the dynamics is fully determined for $i \notin S$ by a set of relations among the activation times of neighboring nodes, which we denote by $t_{i} = \phi_i(\{t_j\})$ with $j \in \dd i$.
In terms of activation times, the dynamical rule for the LTM translates into $t_i=\phi_i(\{t_j\})$ with
\begin{equation}
\phi_i(\{t_j\}) = \min \left\{ t \in \mathcal T : \textstyle{\sum}_{j\in\dd i} w_{ji}\1[t_j<t]\geq \theta_i \right\}.
\end{equation}
Admissible trajectories in this model correspond to vectors $\bt$ such that $\Psi_i = \1 \left[ t_i = 0 \right] + \1\left[t_{i} = \phi_i(\{t_j\}) \right]$ equals $1$ for every $i$.

In this static representation, one can introduce an energetic term $\mathscr E(\bt)$ that gives different probabilistic weights to different trajectories. The path probability associated to a configuration of activation times is
\begin{equation}\label{path}
P(\bt) = \frac{1}{Z} e^{-\beta \mathscr E(\bt)}  \prod_{i\in V}\Psi_i(t_i,\{t_j\}_{j\in \dd i})
\end{equation}
with $Z = \sum_{\bt} e^{-\beta \mathscr E(\bt)}  \prod_{i}\Psi_i(t_i,\{t_j\}_{j\in \dd i})$.
The large deviations properties of the dynamical process can be studied evaluating the static partition function for the dynamic trajectories with an opportunely defined energetic term.
Notice that the value chosen for $T$ will affect the ``speed'' of the propagation: a lower value of $T$ will restrict the optimization to ``faster'' trajectories, at the (possible) expense of the value of the energy.

The most general form of energy function we consider is $\mathscr E(\bt) = \sum_i \mathscr E_i(t_i)$ where $\mathscr E_i(t_i)$ is the ``cost'' (if positive, or ``revenue'' if negative) incurred by activating vertex $i$ at time $t_i$. In the following, we set $\mathscr{E}_i(t_i) =  \mu_i \1 \left[t_i = 0\right] - \epsilon_i \1 \left[ t_i < \infty \right]$ where $\mu_i$ is the cost of selecting vertex $i$ as a seed, and $\epsilon_i$ is the revenue generated by the activation of vertex $i$. Variants with arbitrarily signed parameters $\mu_i,\epsilon_i$ are also possible.
Trajectories with small energy will have a good trade-off between the total cost of their seeds and the total revenue of active nodes. For $\epsilon=0$, the Boltzmann weight reproduces the dynamics of direct propagation from randomly drawn sets of seeds discussed in the previous section. In this case, it can be shown that the equations become equivalent to those presented in \eqref{eq-rho}-\eqref{eq-chi} for direct propagation analysis. On the other hand, in the case $\epsilon> 0$, the causal representation of the dynamics as a DAG is not sufficient to solve the optimization problem as this now implies a backward propagation of information from time $t=\infty$ to time $t=0$.
\begin{figure}
\includegraphics[width=0.6\columnwidth]{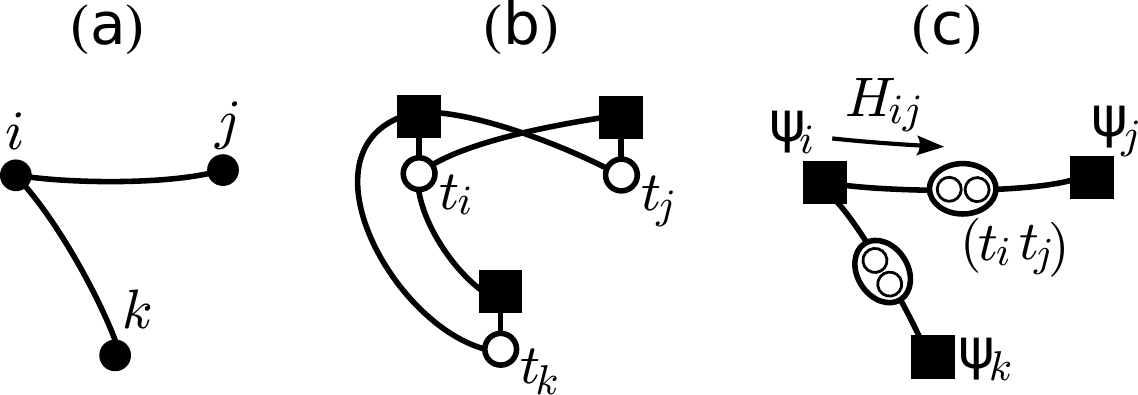}
\caption{Dual factor graph representation for the spread optimization problem. (a) Original graph. (b) Naive factor graph formulation, including small loops. (c) Dual factor graph formulation, with variables nodes $(t_i,t_j)$ and $(t_i,t_k)$ and factor nodes $\Psi_i,\Psi_j,\Psi_k$. Factor $\Psi_i$ must ensure, additionally to the dynamical constraint for vertex $i$, that $t_i$ components of $(t_i,t_j)$ and $(t_i,t_k)$ coincide.}
\label{factor}
\end{figure}

\subsection{Derivation of the BP equations}

The representation of the dynamics as a high dimensional static constraint-satisfaction model over discrete variables (i.e. the activation times) defined on the vertices of a graph makes it possible to apply the cavity method \cite{MP} and to develop efficient message-passing algorithms, such as  Belief-Propagation (BP) and Max-Sum (MS). As usual in combinatorial optimization, the variables and their constraints can be represented by means of a factor graph.
However, in the static representation of the dynamics, every constraint $\Psi_i$ depends on the values of all activation times in the neighborhood of node $i$, therefore nearby constraints
$\Psi_i$ and $\Psi_j$ share the two variables $t_i$ and $t_j$ leading to the appearance of short loops in the corresponding factor graph.
In order to eliminate these systematic short loops, we employ a dual factor graph in which variable nodes representing the pair of times $(t_i, t_j)$ are associated to edges $(i, j) \in E$, while the factor nodes are associated to the vertices $i$ of the original graph $G$ and enforce the hard constraints $\Psi_i$  and the contribution $\mathscr{E}_i$ from $i$ to the energy. Figure \ref{factor} gives an illustrative example of such dual construction.
Whenever the original graph is locally a tree, the dual factor graph is such as well. This property allows one to employ the cavity method.
Since the variables appearing in the dual graph are pairs of times $(t_i, t_j)$, the full distribution can be parametrized in terms of cavity marginals $H_{ij}(t_i, t_j)$ for pairs of times. Let us consider the path probability in \eqref{path} and marginalize over all variables but $j$ to compute the probability $P_j(t_j)$ that node $j$ activates at time $t_j$. On an infinite tree we have
\begin{equation}
P_j(t_j) \propto \sum_{\{t_i\}_{i\in \partial j}} e^{-\beta \mathscr{E}_j(t_j)} \Psi_j(t_j,\{t_i\}) \prod_{i\in\partial j} H_{ij}(t_i,t_j)
\end{equation}
where 
the cavity marginal $H_{ij}(t_i, t_j)$ denotes the probability that nodes $i$ and $j$ activate at times $t_i$ and $t_j$ in absence of the constraint $\Psi_j$ and energetic term $\mathscr{E}_j$. It satisfies the recursive relation
\begin{equation}\label{bp}
H_{ij}(t_i, t_{j}) \propto e^{-\beta\mathscr{E}_i(t_i)}\sum_{\{t_k\}}\Psi_i(t_i,\{t_k\})\prod_{k} H_{ki}(t_k , t_i)
\end{equation}
where $k \in \dd i \sm j$. On a general graph, \eqref{bp} define the Belief Propagation (BP) equations that are valid under the hypothesis of fast decay of correlations with the distance or replica symmetric (RS) assumption \cite{MP}.
Under this assumption,  the statistical properties of the system are described by a unique Gibbs state (i.e. replica symmetry), and the BP equations admit a unique solution.

Given a solution of \eqref{bp}, the marginal probability that neighboring nodes $i$ and $j$ activate at times $t_i$ and $t_j$ is $P_{ij}(t_i, t_j)\propto H_{ij}(t_i,t_j) H_{ji}(t_j, t_i)$. Equations \eqref{bp} allow one to access the statistics of atypical dynamical trajectories (e.g. entropies of trajectories or distribution of activation times), but it involves a number of terms which is exponential in the vertex degree. An equivalent but tractable expression can be obtained as follows. For $0<t_{i}<\infty$, Eq.~\eqref{bp} can be expressed as
\begin{equation}
H_{ij}(t_i,t_j) \propto e^{-\beta\mathscr{E}_i(t_i)}\sum_{\substack{\theta_1\geq \theta_i - w_{ji}\1[t_{j}\leq t_{i}-1]\\
\theta_2<\theta_i-w_{ji}\1[t_{j}<t_{i}-1]
}}Q_{ij}^{t_i}(\theta_1,\theta_2)
\end{equation}
where $Q_{ij}^{t_i}$ is the two dimensional convolution of functions $f^{t_i}_{k}(\theta_1,\theta_2)=\sum_{t_{k}}\delta(\theta_1,w_{ki}\1[t_{k}<t_{i}-1])\delta(\theta_2,w_{ki}\1[t_{k}\leq t_{i}-1])H_{ki}(t_{k},t_{i})$ for $k\in \partial i\setminus j$.

In the limit $\beta \to \infty$, with a proper rescaling of the messages, \eqref{bp} gives the Max-Sum (MS) equations and algorithm, which can be used to find explicit solutions at minimum energy.  In this limit, the optimization of the dynamics of the LTM correspond to the {\em spread maximization problem}, which is computationally hard even to approximate in the worst case \cite{LZWKF}.

\begin{figure}[t]
\includegraphics[width=0.6\columnwidth]{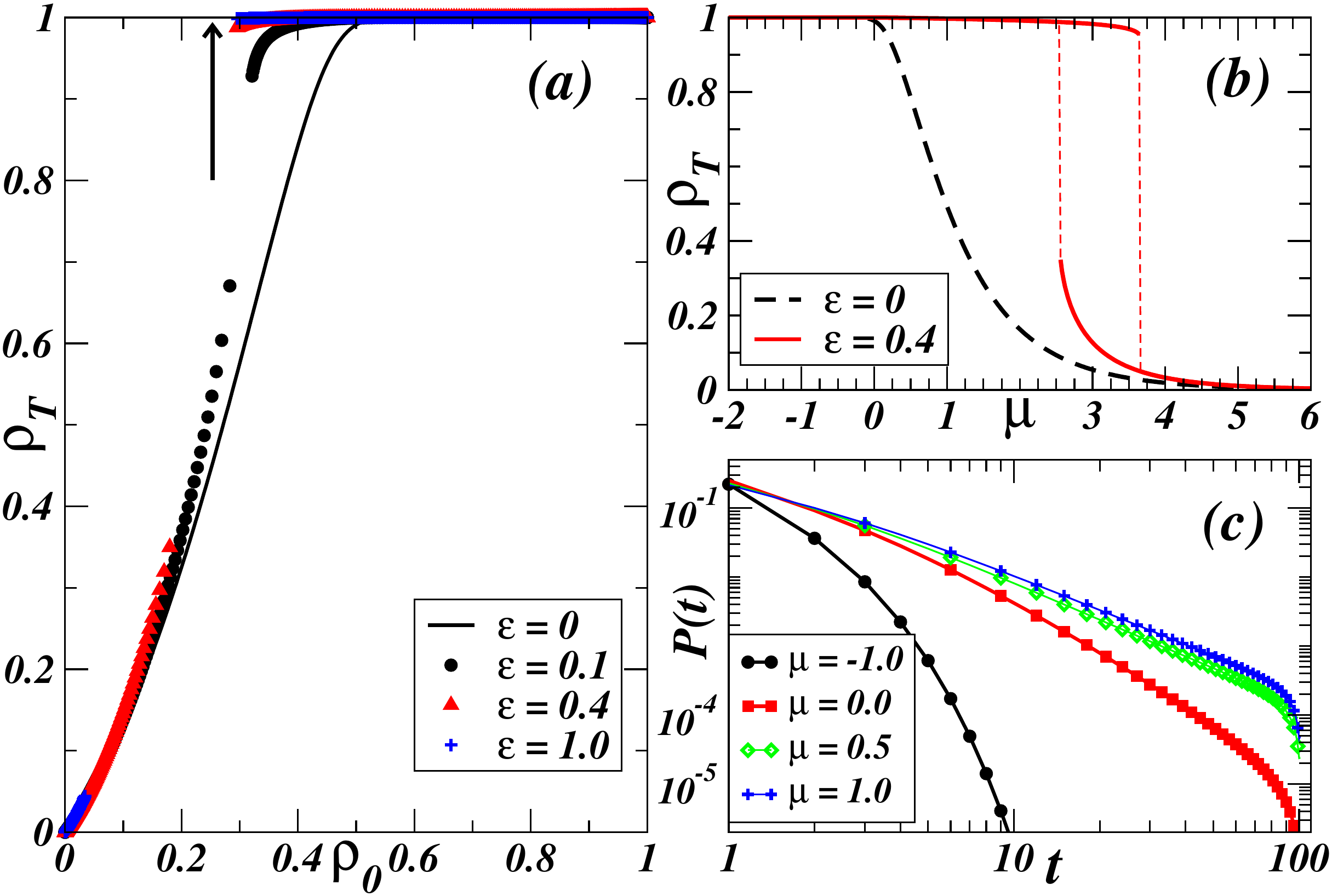}
\caption{(Color online) (a) Parametric plot $\rho_T$ v.s.$\rho_0$ obtained solving the Belief-Propagation (BP) equations in the single-link approximation on regular random graphs of degree $K=3$, for threshold $\theta = 2$, duration $T=20$ and $\epsilon = 0,0.1,0.4,1$. The vertical arrow indicates the minimum density of seeds ($\rho_0\approx 0.253$) necessary for the total activation obtained by the Max-Sum algorithm on finite graphs of size $|V|=10,000$.
(b) Curves $\rho_T(\mu)$ for $\epsilon = 0$ (black dashed line) and $0.4$ (red full line). The latter are obtained following the upper and lower branches of solution across the transition. (c) Activation time probability $P(t)$ obtained computing the total BP marginals in a dynamics of duration $T=100$, for $\epsilon = 0.4$ and different values of $\mu$.
\label{figRRGK3t2}}
\end{figure}

\section{Results on ensembles of random graphs}

On ensembles of (infinitely large) random graphs, the solution of the BP equations \eqref{bp} can be computed at any finite $\beta$ using a population dynamics method in the single-link approximation \cite{MP}.

\subsection{Homogeneous solution on random regular graphs}
For random regular graphs and considering a completely homogeneous setup (i.e.  uniform weights $w_{ij} = 1$ $\forall (i,j) \in E$, uniform thresholds $\theta_i = \theta$, $\forall i \in V$, uniform costs $\mu_i =\mu$, $\forall i \in V$ and uniform revenues $\epsilon_i=\epsilon$, $\forall i \in V$), the replica symmetric cavity marginals are expected to be uniform, therefore the population dynamics can be replaced by a self-consistent equation for a single representative BP marginal $H(t,s)$. Since all incoming links are assumed to have the same set of messages, one can group equal messages together introducing a multinomial distribution and obtaining the following system of nonlinear equations:
\begin{subequations}\label{BPsl}
\begin{align}\label{BPsl1}
H(0,s) & \propto  e^{-\beta \mu} p_0^{K-1} \\
\label{BPsl2} H(t,s) & \propto  \sum_{\substack{n_- + n_+ + n_0 = K-1 \\ n_- < \theta - \1[s<t-1] \\ \theta-\1[s\leq t-1] \leq n_- + n_0}}  \frac{(K-1)!}{n_{-}! n_{+}! n_{0}!} p_t^{K-1-n_-n_0} m_t^{n_-} H(t-1,t)^{n_0} \quad \quad \text{for} \quad 0 < t \leq T \\
\label{BPsl3} H(\infty,s) & \propto  e^{-\beta \epsilon} \sum_{n_{-} \leq \theta -1 -\1[s<T]} \binom{K-1}{n_{-}} \left[ H(T,\infty) + H(\infty, \infty) \right]^{K-1-n_{-}} m_{\infty}^{n_{-}}
\end{align}
\end{subequations}
where we defined the cumulative messages $p_{t} = \sum_{t'\geq t} H(t',t)$ and $m_t = \sum_{t' < t-1} H(t',t)$. The normalization constant is just the sum of all messages. The system of equations could be further simplified from $O(T^2)$ messages to $O(T)$ by exploiting the fact that $H(t,s) = H(t,\sign(t-s+1))$.

The behavior of \eqref{BPsl} can be studied varying $\mu,\epsilon,\beta$ and $T$ for any given assignment of $K$ and $\theta$. We consider the representative cases $K=3, \theta=2$ and $K=4, \theta =2$ in the $(\epsilon,\mu)$-plane at fixed $T$ and $\beta=1$, then we will comment on the effects of varying $T$ and $\beta$. As for the direct dynamics in Sec.\ref{sec2b}, we shall consider as observables the density of seeds $\rho_0$ and the final density (at time $T$) of active nodes $\rho_T$. For $\epsilon=0$ and $T\to \infty$, we recover known results for the static of Bootstrap Percolation \cite{CLR79,DGM06,BP07} in Sec.\ref{sec2b}. Although for finite $T$ both cases present a continuous behavior at $\epsilon=0$, the two activation mechanisms are qualitatively different and this difference is amplified in the large deviations regime.

\begin{figure}[t]
\includegraphics[width=0.6\columnwidth]{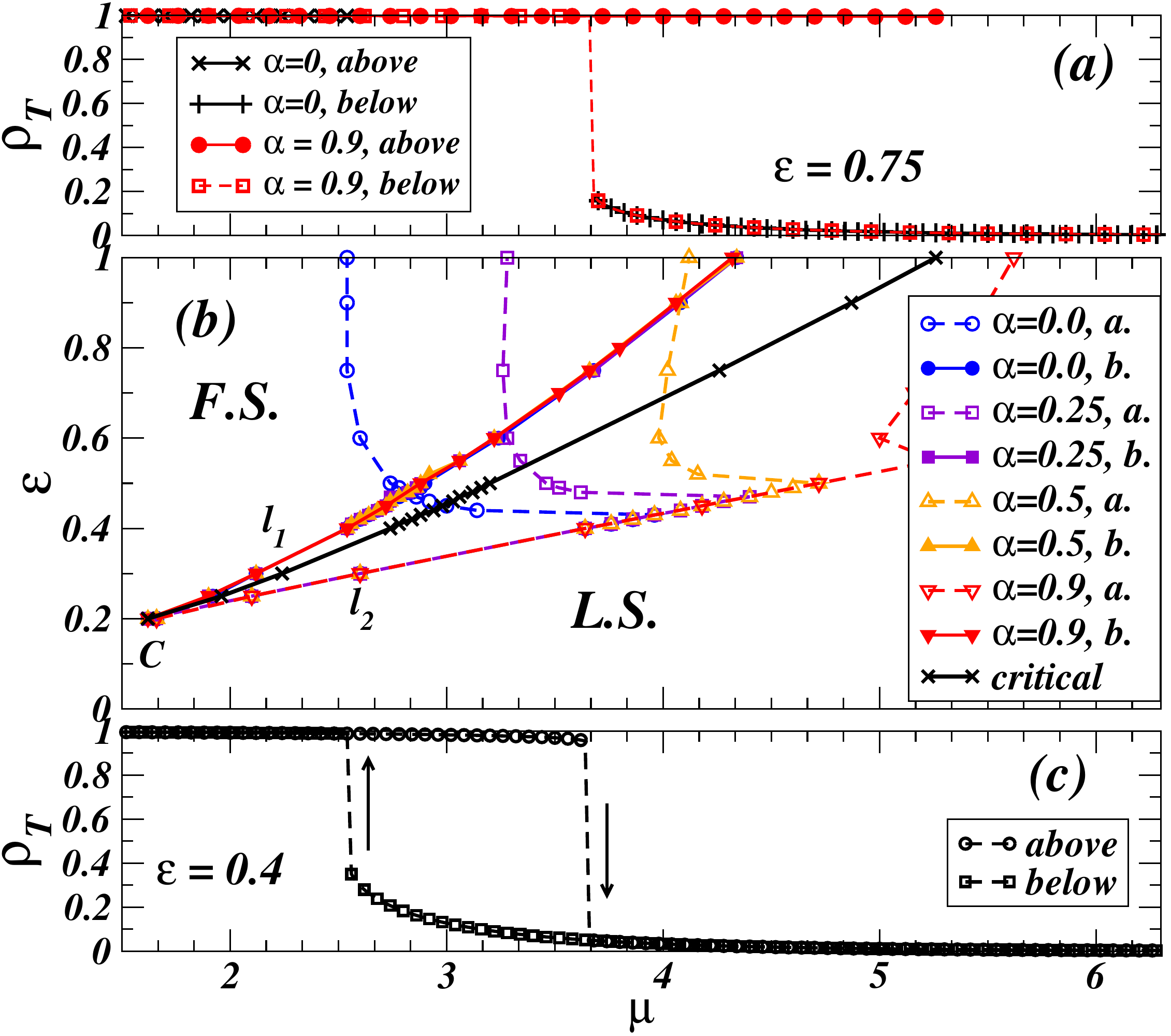}
\caption{(Color online) (b) Phase diagram as a function of $\epsilon$ and $\mu$ for $T=20$ in random regular graphs of degree $K=3$ and thresholds $\theta=2$. The curves are obtained in the single-link approximation using a damping procedure to improve the convergence. Black crosses indicate the location of the thermodynamic phase transition (crossing of the free-energy branches) between a full spread (F.S.) phase and a low spread (L.S.) phase. The upper and lower curves, $\ell_1$ and $\ell_2$, in the same plot indicate the spinodal lines, obtained for increasing values of the damping factor $\alpha$ from 0 to 0.9.  The panel (c) shows the hysteresis phenomenon observed in the density of activated nodes $\rho_T$ as function of the chemical potential $\mu$ for $\epsilon = 0.4$. The same phenomenon for $\epsilon = 0.75$ is shown in panel (a), where we see that the lack of convergence for $\alpha=0$ is cured by improving the damping to $\alpha = 0.9$.
}\label{epsmuT20K3t2}
\end{figure}

\subsection{Case $K=3, \theta =2$}
For random initial conditions ($\epsilon=0$), the density $\rho_T$ of active nodes in the final state is a continuous function of $\rho_0$. Fig.\ref{figRRGK3t2}a shows that under optimization the curves develop a gap in the possible values of $\rho_0$ and $\rho_T$ obtained by varying $\mu$. This means that (for sufficiently large $\epsilon$ and $\beta$) a value $\mu^*$ exists at which both $\rho_0$ and $\rho_T$ undergo a discontinuous transition, with coexistence and hysteresis phenomena (see Fig.\ref{figRRGK3t2}b). As $\beta$ increases the minimum density of seeds admitting full spread ($\rho_T=1$) gradually approaches the values obtained by the MS algorithm (zero-temperature limit of the BP equations). The total marginal computed from \eqref{BPsl} gives the probability $P(t)$ that a node gets activated at time $t$. The activation time distribution $P(t)$ is displayed in Fig.\ref{figRRGK3t2}c for $T=100$. While for $\epsilon=0$ it always decays exponentially, for $\epsilon>0$ it develops a power-law shape when $\mu$ is increased towards the region in which optimization is effective. It means that in order to optimize the dynamics, one can decrease the number of seeds at the cost of generating an activation process that proceeds at a slower pace. The longer the allowed duration $T$, the smaller the minimum density of seeds required to reach full spread under optimization, but the larger the tail of the distribution.

A tentative phase-diagram in the $(\mu,\epsilon)$-plane, corresponding to the solution of \eqref{BPsl} with $T=20, \beta=1$, is displayed in Fig.\ref{epsmuT20K3t2}. The results are only partially correct because the BP equations do not converge for all values of the parameters. Increasing $\epsilon$ from 0, the transition is still continuous, until we encounter a tricritical point $C = (\mu^*, \epsilon^*)$ where the activation transition becomes discontinuous with the appearance of a coexistence phase that grows with $\epsilon >\epsilon^* \approx 0.2$. For moderately small values of $\epsilon$ (e.g. $\epsilon = 0.4$ in the bottom panel of Fig.\ref{epsmuT20K3t2}),
the BP equations converge to their fixed-points and the behavior of the system can be correctly studied for all values of $\mu$. We used a cooling/annealing scheme in $\mu$ at fixed $\epsilon$, that allowed us to follow the upper (high $\rho_T$, low $\mu$) and lower (low $\rho_T$, high $\mu$) branches of the curve $\rho_T(\mu)$ even into the coexistence region. The coexistence phase is limited by two spinodal lines $l_{1}$ and $l_{2}$ departing from  $(\mu^*, \epsilon^*)$ and indicating the location where the two branches of solutions terminate. It is possible to locate the discontinuous phase transition by comparing the free-energy of the two solutions in the coexistence region (black crosses). The meaning of the spinodal lines becomes evident looking at the bottom panel of Fig.\ref{epsmuT20K3t2}, in which we show the behavior of the solutions across the coexistence region for $\epsilon = 0.4$ (see also Fig.\ref{figRRGK3t2}B).

Surprisingly, the spinodal line $l_2$ (open symbols) seems to present a non-monotonic behavior with $\epsilon$. This result is just a non-physical artifact of the lack of convergence of the iteration procedure used to compute the fixed-points of \eqref{BPsl}. In order to improve convergence also for large values of $\epsilon$, we used a ``damped" update rule, in which at each iteration, every message is replaced by a linear combination of her old and new values, i.e. $H^{old}(t,s) \leftarrow \alpha H^{old}(t,s) + (1-\alpha) H^{new}(t,s)$ with $\alpha \in [0,1]$.
Increasing the damping factor $\alpha$, the convergence properties of \eqref{BPsl} are improved and the line $l_{2}$ correctly moves smoothly towards larger values of $\mu$ (red line with downward triangles points).
In the central panel of Fig.\ref{epsmuT20K3t2} we show the effect of non-convergence on the curves $\rho_T(\mu)$ for $\alpha = 0, 0.25, 0.5, 0.9$. The top panel reports the same plot of the bottom one, namely $\rho_T(\mu)$, for a larger value of $\epsilon$ where the BP equations do not converge without damping. The improvement obtained with a damping factor $\alpha = 0.9$ is evident.

A more sophisticated way of stabilizing the solution scheme for the BP equations \eqref{BPsl} is that of using a population of $\mathcal{N}_{pop} \gg 1$ messages. Solving the \eqref{BPsl} using population dynamics is very time consuming, but on RRGs the results with $\mathcal{N}_{pop} \simeq 10^3\div10^5$ are in agreement with the results obtained using the damped BP equations.

\begin{figure}[t]
\includegraphics[width=0.6\columnwidth]{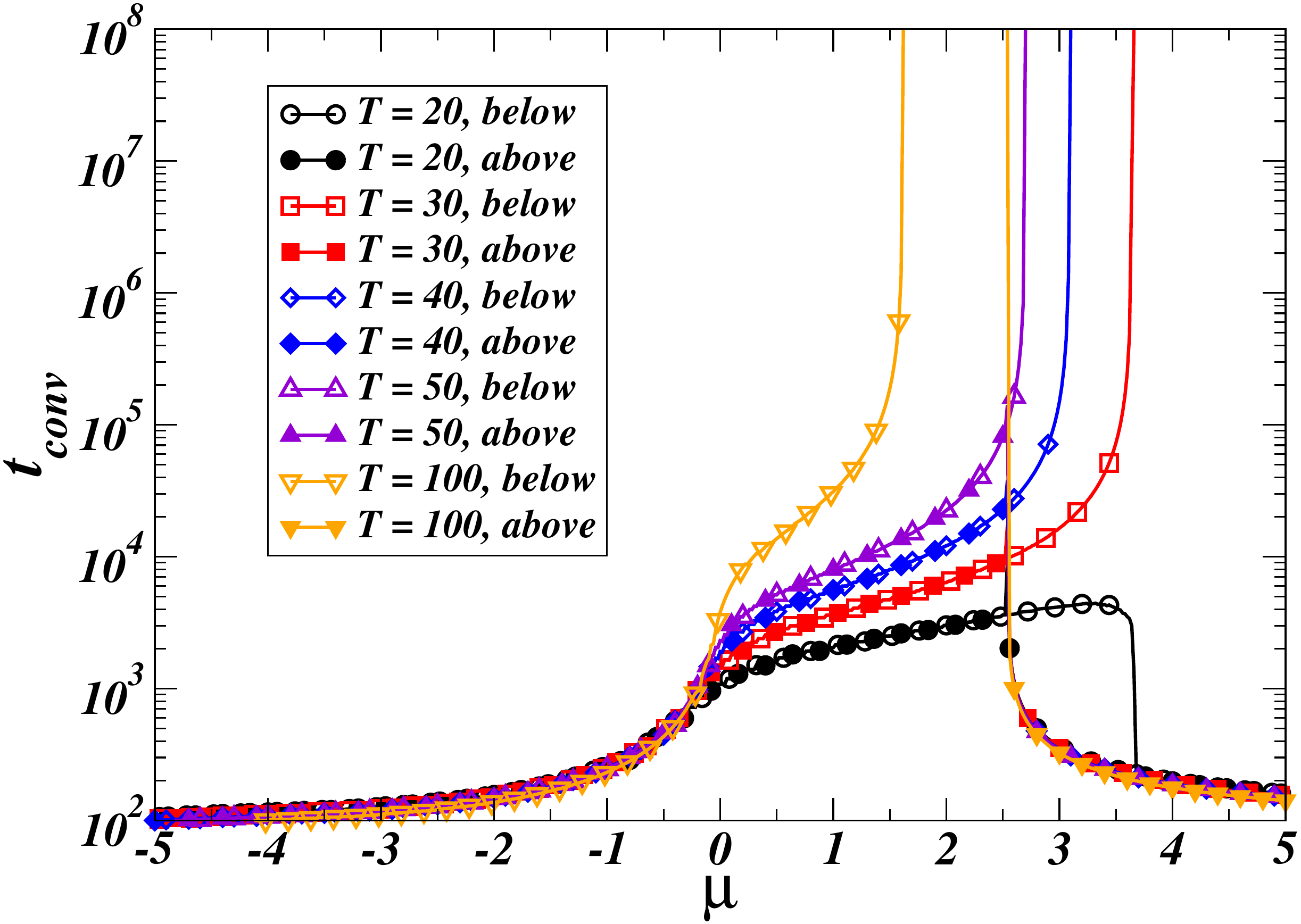}
\caption{(Color online) Convergence time (number of iterations) of BP equations for $K=3$, $\theta=2$ in the single-link approximation at $\epsilon = 0.4$ as a function of $\mu$ and damping factor $\alpha=0.9$.
Different symbols and colors correspond to different values $T$ of the length of the dynamics. At each value of $T$ we reported two different curves (open and full symbols) corresponding to experiments performed increasing or decreasing the values of $\mu$ (in this way following the two branches of solutions).
}\label{tconvK3t2}
\end{figure}

\begin{figure}
\includegraphics[width=0.6\columnwidth]{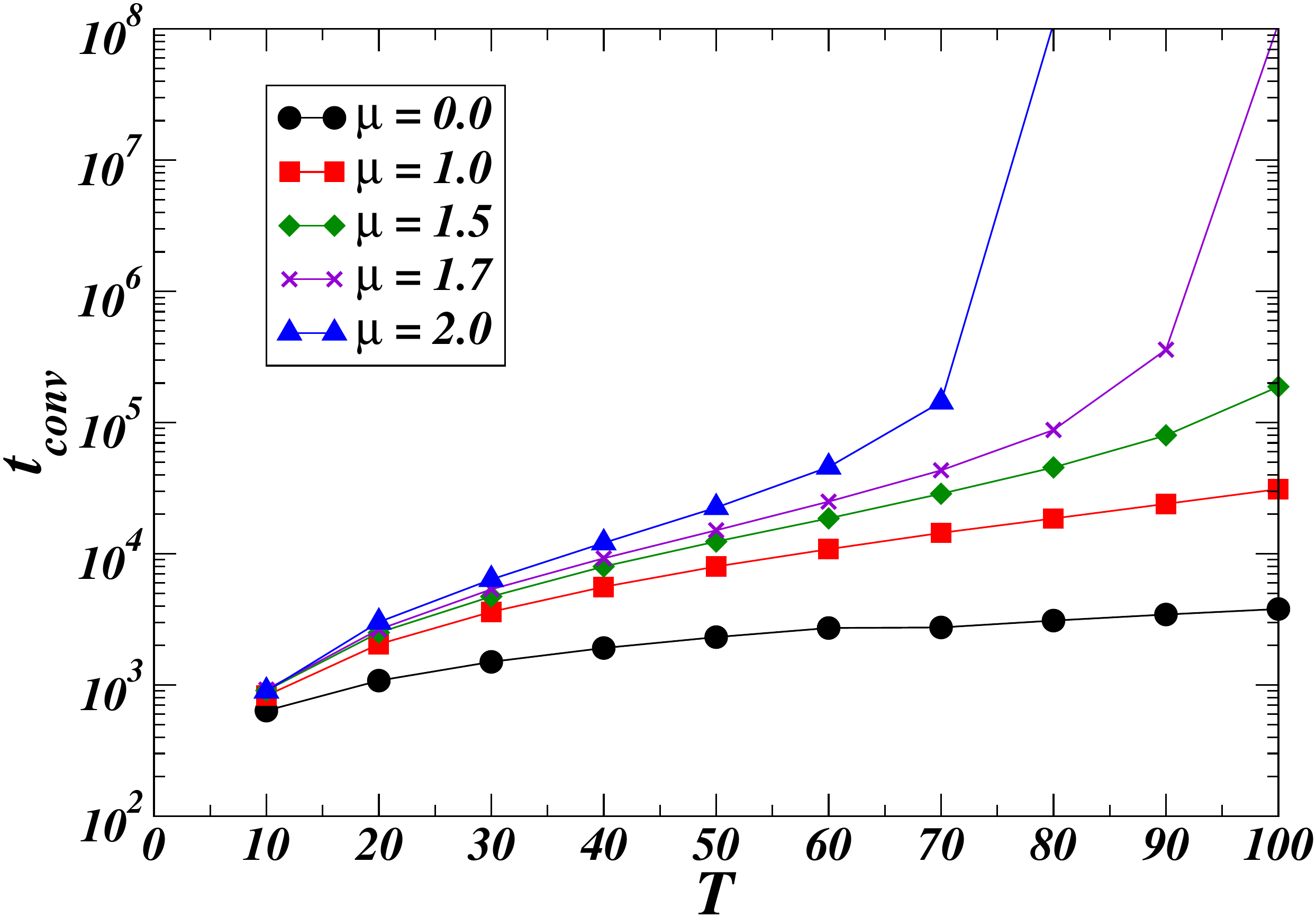}
\caption{(Color online) Convergence time of the BP equations as function of $T$ for different values of $\mu$ in random regular graphs with $K=3$, $\theta=2$ and $\epsilon = 0.4$.
}\label{scalingT}
\end{figure}

Fig.\ref{tconvK3t2} displays the number of iterations $t_{conv}$ necessary to reach the fixed point of \eqref{BPsl} as a function of $\mu$ for $\epsilon = 0.4$ with damping factor $\alpha = 0.9$. In the region where the optimization of the dynamical process is effective, the convergence time grows continuously until it diverges. For small values of $T$ (e.g. $T=20$) there is no divergence. At larger $T$ the BP equations stop converging at values of $\mu$ that decrease with increasing $T$. The dependence on $T$ at fixed values of $\mu$ is reported in Fig.\ref{scalingT} for $\epsilon=0.4$. On the contrary, the lack of convergence appears abruptly at the spinodal line (independently of the damping factor $\alpha$) when decreasing $\mu$ from large positive values (Fig.\ref{tconvK3t2}).

\begin{figure}[t]
\includegraphics[width=0.6\columnwidth]{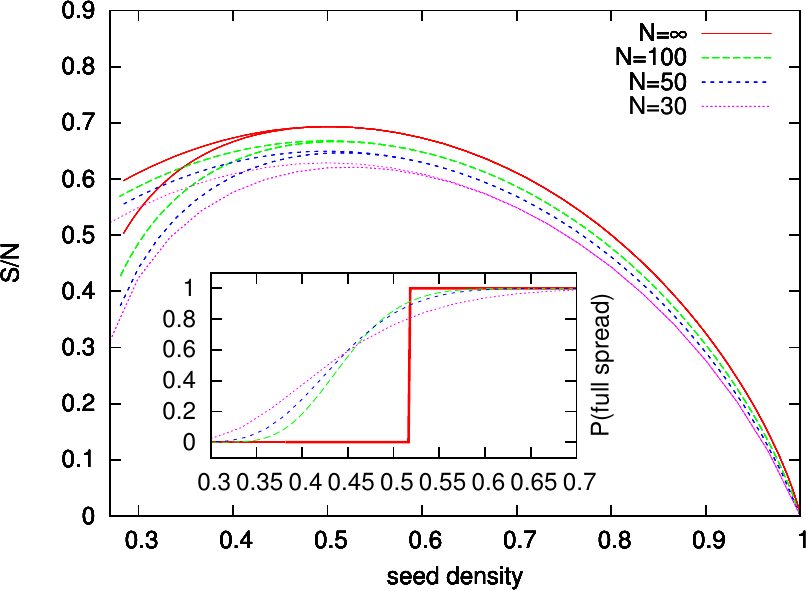}
\caption{(Color online) Entropy per site $S/N$ of the solutions of the full spread problem on regular random graphs of degree $K=3$ and threshold $\theta = 2$, for $N=30$, $50$, $100$, $\infty$ vs. seed density $\rho_0$. For each $N$ the upper line corresponds to the normalized binomial distribution (i.e. per site entropy of seeds in the absence of optimization) and the lower one to the entropy per site of fully spreading seeds. Inset: Probability $P$ of randomly selecting a fully spreading set of seeds for the same set of parameters.
}\label{entropy}
\end{figure}

The above results show that, when the infinite time limit of the direct dynamics presents a continuous activation transition, the optimization of the spreading process is possible and effective. One could also verify how rare the optimal trajectories are by computing their entropy and comparing it with the entropy of random trajectories. We did it in Fig.\ref{entropy} where we plot the entropy $s$ of the initial conditions that lead to a full spread as function of the density of seeds. The result for $N=30$ is obtained by explicit enumeration, whereas for larger systems ($N=50,100$) we used a generalization of the cavity method that allows to fix a global constraint (the number of seeds) by introducing an additional set of messages that flow over a spanning tree superimposed on the original graph (see \cite{BRZZ11} Appendix B). The curves for the limit of infinite random regular graphs are obtained computing the entropy, in the cavity approximation, from the fixed-point solution of \eqref{BPsl}.
These quantities for non-typical trajectories are compared with the entropy curves associated to a random choice of initial conditions with fixed density of active nodes $\rho_0$, that is given by a binomial sampling of initial seed nodes. When the curves deviate from the binomial, the probability of choosing randomly an optimal set of seeds becomes exponentially small (inset in Fig.\ref{entropy}). In the infinite system, this event is governed by a zero-one law.

\begin{figure}[t]
\includegraphics[width=0.6\columnwidth]{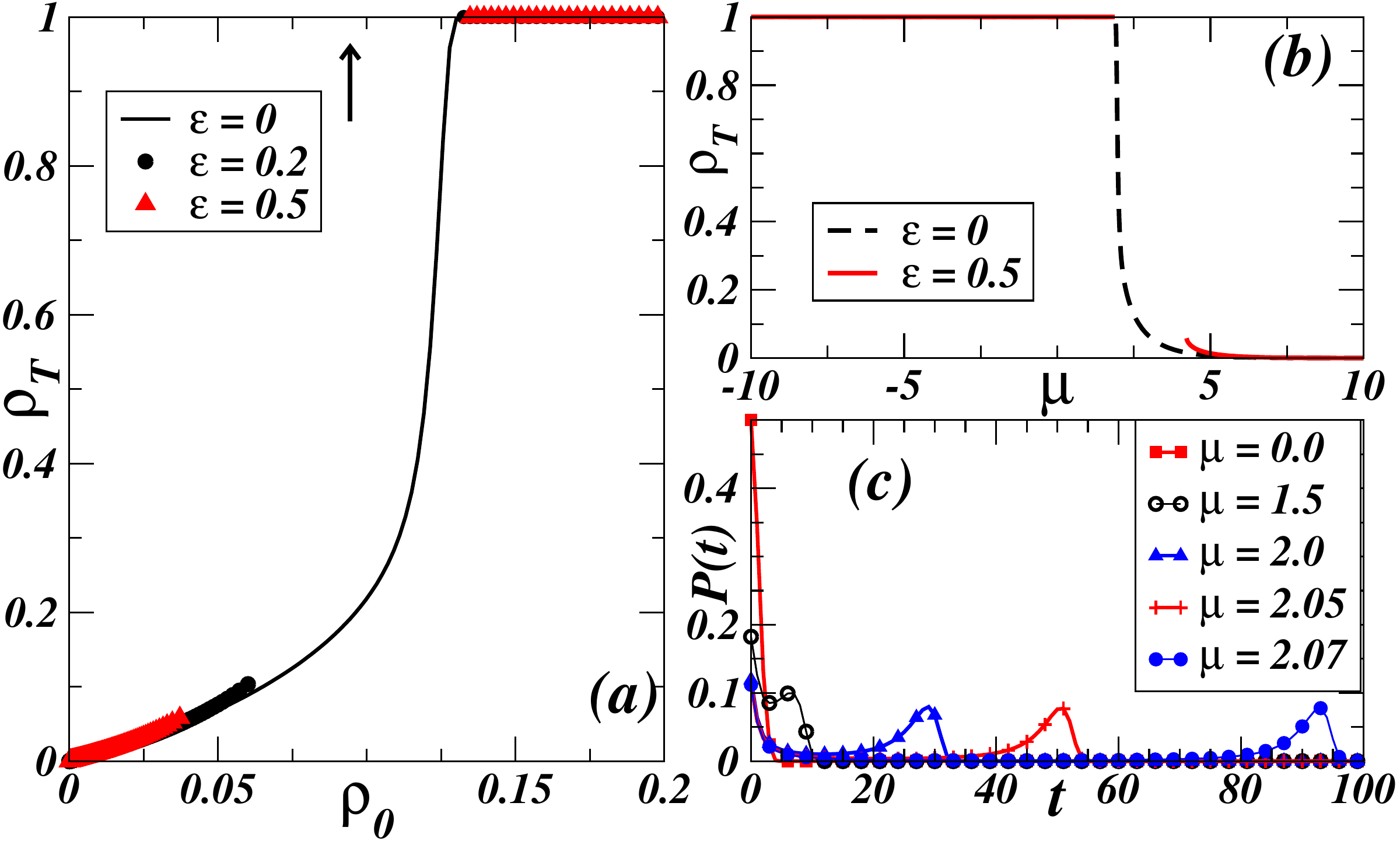}
\caption{(Color online) (a) Parametric plot $\rho_T$ v.s.$\rho_0$ obtained solving \eqref{BPsl} in the single-link approximation on regular random graphs of degree $K=4$, for threshold $\theta = 2$, duration $T=20$ and $\epsilon = 0,0.2,0.5$. No improvement respect to the random case ($\epsilon=0$) is obtained by performing optimization ($\epsilon = 0.2,0.5$). The vertical arrow indicates the minimum density of seeds ($\rho_0\approx 0.094$) necessary for the total activation obtained by the Max-Sum algorithm on finite graphs of size $|V|=10,000$.
(b) Curves $\rho_T(\mu)$ for $\epsilon = 0$ (black dashed line) and $0.5$ (red full line). For $\epsilon = 0.5$ the BP equations do not converge in a region of values of $\mu$ close to the discontinuous transition. (c) Activation time probability $P(t)$ obtained computing the total BP marginals in a dynamics of duration $T=100$, for $\epsilon = 0$ and different values of $\mu$ (for $\epsilon = 0.5$ we obtain exactly the same behavior).
}\label{figRRGK4t2}
\end{figure}

\subsection{Case $K=4, \theta =2$}

In this case, the discontinuous behavior is already present at $\epsilon=0$ in the limit of large $T$. For $T=20$ the curve $\rho_T$ vs. $\rho_0$ in Fig.\ref{figRRGK4t2}  is very steep but continuous, however the underlying dynamics is qualitatively different from that of $K=3, \theta=2$ as discussed in Sec.\ref{sec2b}. The plot of $\rho_T$ vs. $\rho_0$ in Fig.\ref{figRRGK4t2}a shows that some almost no improvement in the density of activated nodes $\rho_T$ is obtained by increasing $\epsilon>0$. Indeed, for $\epsilon>0$, the BP equations converge as long as $\mu$ is smaller than the critical value corresponding to the abrupt transition for $\epsilon =0$, then they stop converging in the region where optimization is expected to be effective (see also Fig.\ref{figRRGK4t2}b). Remarkably, the MS algorithm (supplemented by a reinforcement method \cite{BBBCDFZ10}) finds full-spread solutions that are considerably better than best BP results (arrow in Fig.\ref{figRRGK4t2}).

The activation probability $P(t)$ is very different from the previous case. For random seeds ($\epsilon = 0$) the shape of $P(t)$ is not monotonically decreasing, but it develops a second peak that moves towards large times when $\rho_0$ approaches (from above) the critical value corresponding to the abrupt activation transition. At finite $T$, the existence of such a peak is a precursor of the discontinuous transition that only occurs for $T\to \infty$.
In this limit, the position of the second peak diverges as $\rho_0$ approaches the critical point from above. Figure \ref{figRRGK4t2}c shows the behavior of $P(t)$ for $\epsilon =0$. As long as the BP equations converge, for $\epsilon > 0$ the behavior is the same as that observed for $\epsilon=0$, with a second peak that appears and gradually moves towards larger times when increasing $\mu$ and approaching the threshold of full activation. The peak identify a  ``critical mass'' of nodes  whose dynamical properties are strongly correlated and that activate almost at the same time. The fact that $P(t)$ does not change for $\epsilon >0$ is a clue that a large fraction of variables are strongly correlated already at $\epsilon =0$. The lack of convergence could be due to the onset of long-range correlations responsible of the abrupt activation for $\epsilon=0$ (and $T=\infty$). It also suggests that there are regions of the parameters in which we expect the space of trajectories to display a complex geometrical structure (e.g. clustering phenomena \cite{clustering,clustering2}) that cannot be captured by the simple ``replica symmetric'' \cite{MP} cavity assumption employed here.

Since the effect of the optimization is that of selecting trajectories that allow to postpone as much as possible this sudden activation, the possibility of controlling the trade-off between the total propagation time $T$ and the number of seeds $\rho_0$ required to achieve a certain $\rho_T$ is a potentially useful feature of the proposed message-passing algorithms.

\begin{figure}
\includegraphics[width=0.6\columnwidth]{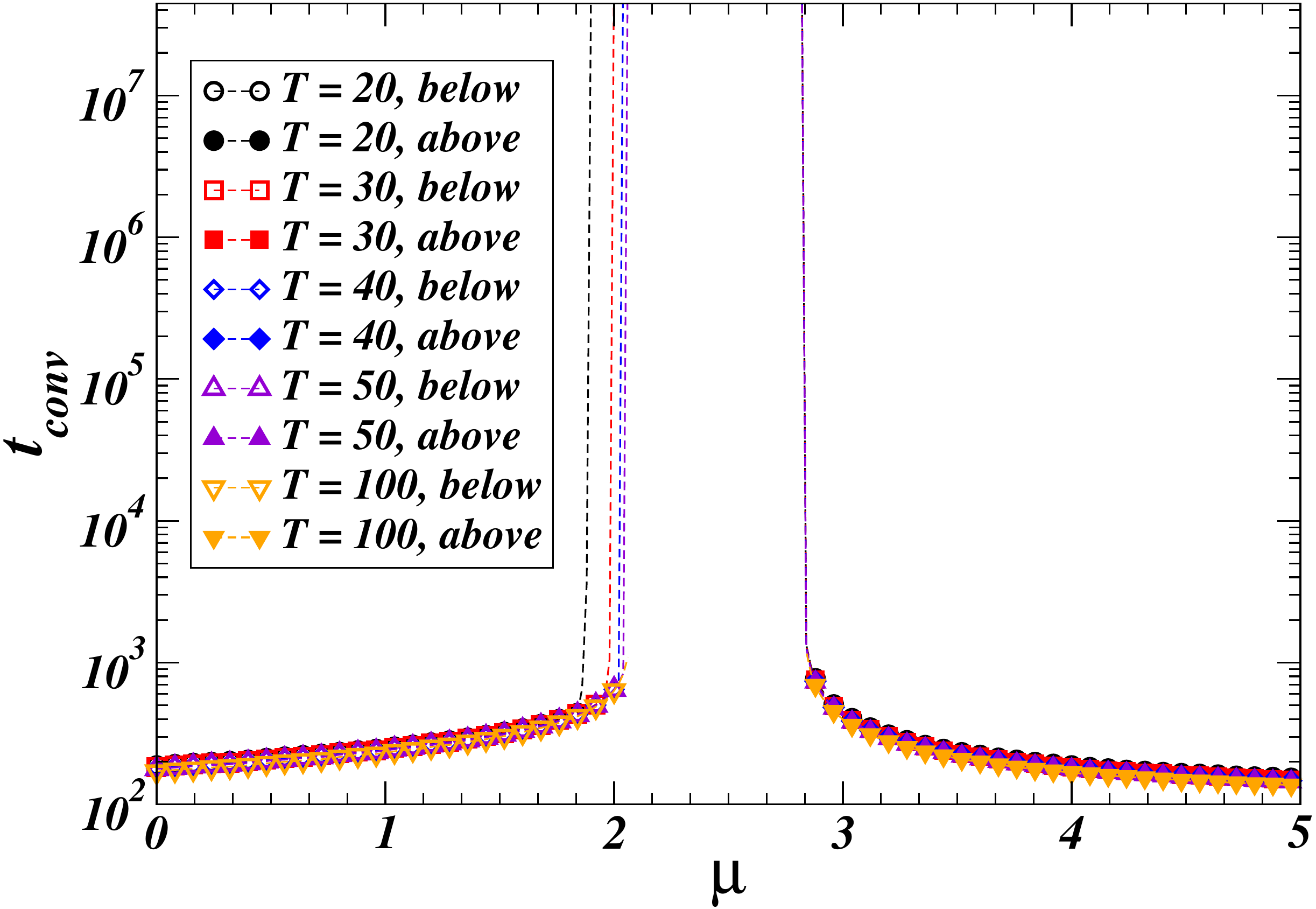}
\caption{(Color online) Convergence time (number of iterations) of BP equations for  $K=4$, $\theta=2$ in the single-link approximation at $\epsilon = 0.5$ as a function of $\mu$ for different values of $T$. 
}\label{timeK4t2eps01}
\end{figure}

\begin{figure}[t]
\includegraphics[width=0.6\columnwidth]{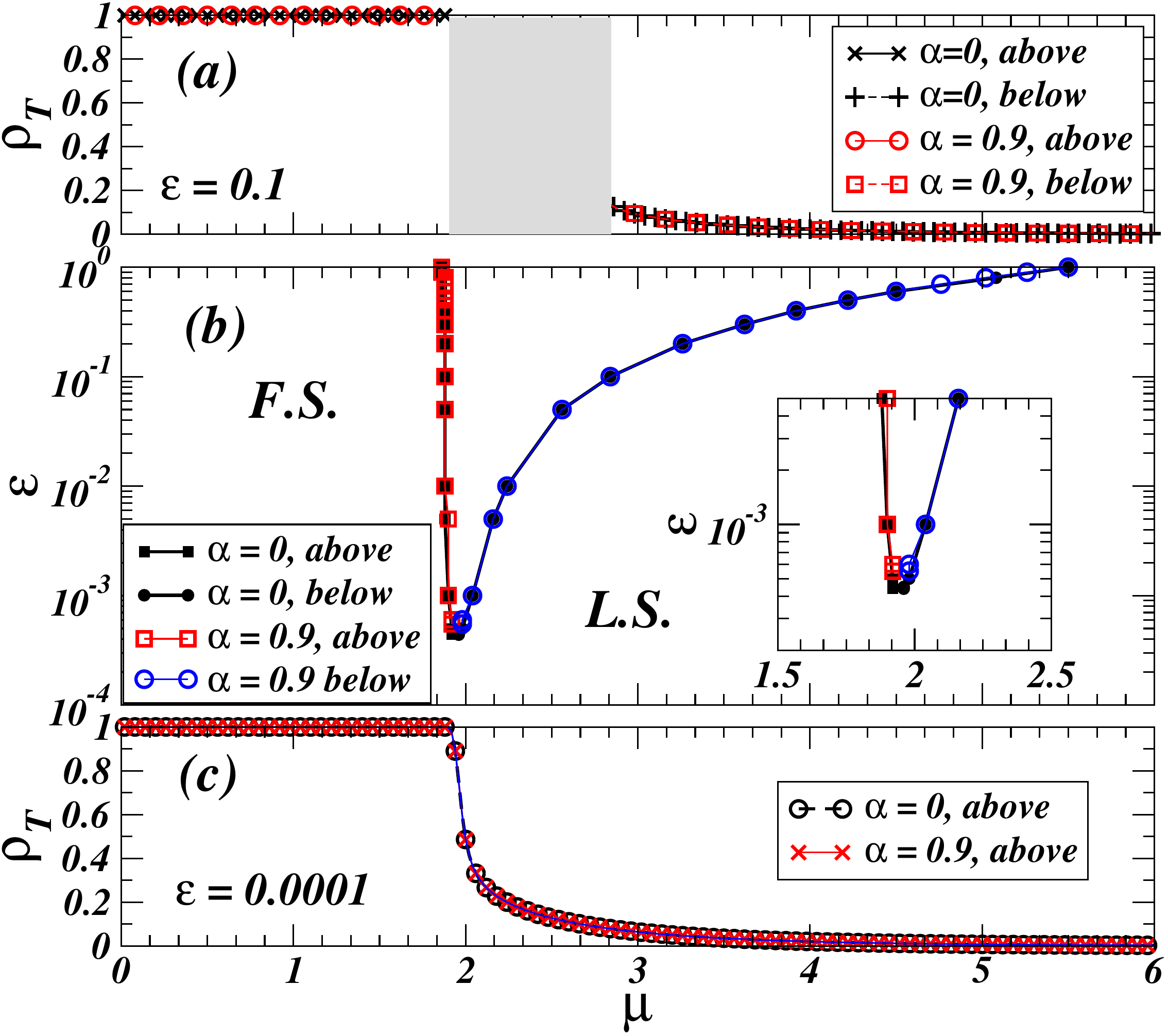}
\caption{(Color online) (b) Phase diagram as a function of $\epsilon$ and $\mu$ for $T=20$ in random regular graphs of degree $K=4$ and thresholds $\theta=2$. Full spread (F.S.) and low spread (L.S.) regions are marked. The curves represent the lines at which BP stops converging in the single-link approximation. The damping procedure (results for $\alpha = 0.9$) does not improve the convergence. The inset highlights the behavior for very small values of $\epsilon$. (c) In this region we can still observe the BP equations converge at some non-zero $\epsilon$. (a) The behavior of the density of activated nodes $\rho_T$ as function of the chemical potential $\mu$ for larger values of $\epsilon$ where the BP equations stop converging in a finite interval of values of $\mu$ (shaded area).
}\label{epsmuT20K4t2}
\end{figure}

The fact that the lack of convergence of the single-link BP equations is here very different from that observed in the case of continuous activation processes can be understood also from the plot of the convergence time of the single-link BP equations in Figure \ref{timeK4t2eps01} that turns out to be almost independent of $T$. The non-convergence persists if we consider a population of $\mathcal{N}_{pop} \gg 1$ messages in the single-link approximation and, for the same range of parameter values, it occurs also when solving BP equations on single instances of graphs.
In Figure \ref{epsmuT20K4t2}, we show the phase diagram in the $(\mu,\epsilon)$-plane for $T=20, \beta=1$. Already at very small values of $\epsilon$  the BP equations stop converging in a region of values of $\mu$ that grows with $\epsilon$. Open symbols show that no improvement is obtained using the damping procedure.

\subsection{Erd\H{o}s-R\'enyi random graphs}
\begin{figure}
\includegraphics[width=0.6\columnwidth]{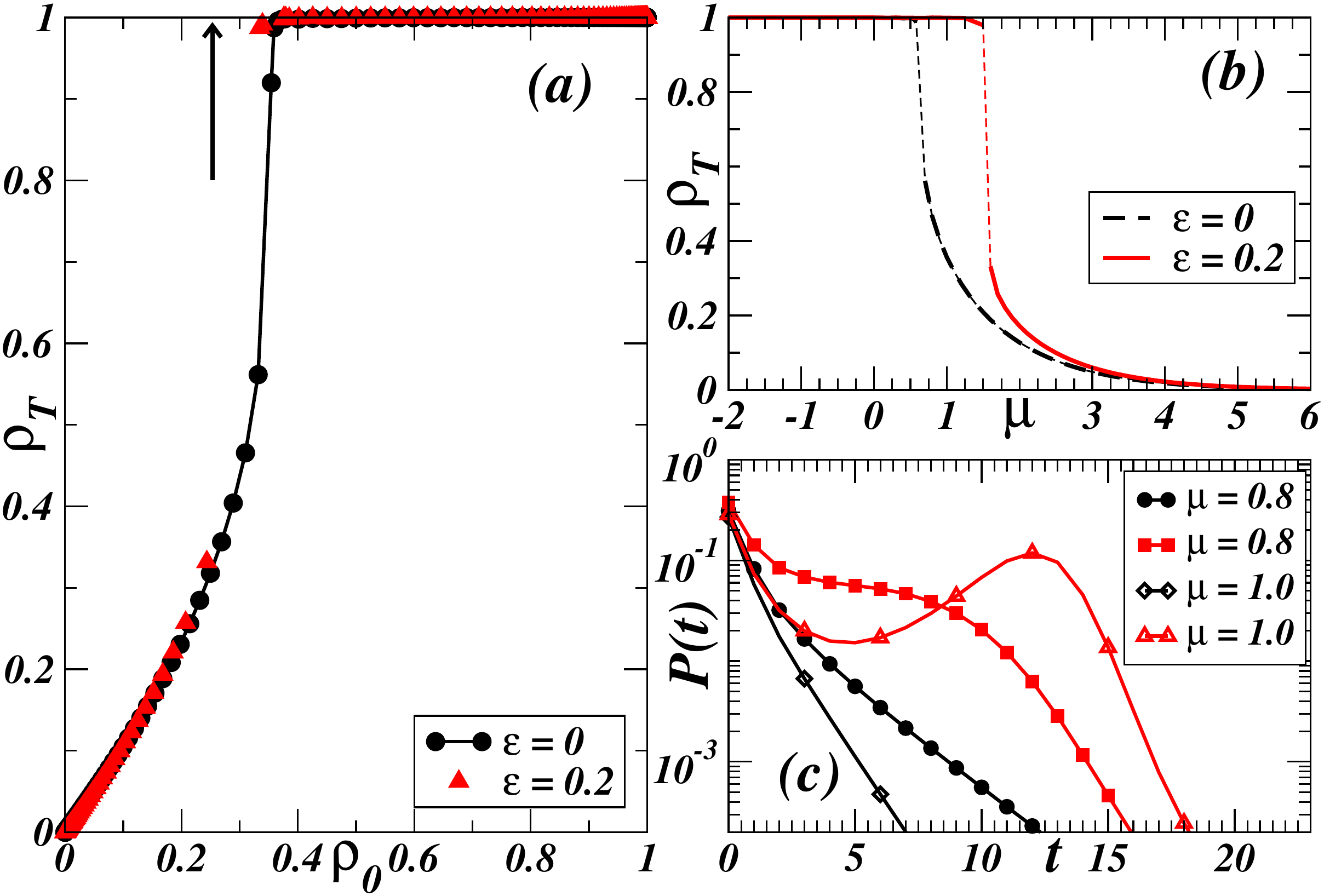}
\caption{(Color online) (a) Parametric plot $\rho_T$ v.s.$\rho_0$ obtained solving the BP equations in the single-link approximation on Erd\H{o}s-R\'enyi random graphs of average degree $z=5$ (minimum degree $k_{min} =1$ and maximum degree $k_{max}= 16$), with thresholds $\theta_i = \lfloor (k_i  +1)/2 \rfloor$, duration $T=20$ and $\epsilon = 0,0.2$.
(b) Curves $\rho_T(\mu)$ for $\epsilon = 0$ (black dashed line) and $0.4$ (red full lines). (c) Activation time probability $P(t)$ obtained computing the total BP marginals in a dynamics of duration $T=20$, for $\epsilon = 0$ (black circles and diamonds) and $0.2$ (red squares and triangles) and different values of $\mu=0.8,1.0$.
\label{figRGz5}}
\end{figure}

To relax the assumption of complete homogeneity of the graphs, we also considered Erd\H{o}s-R\'enyi (ER) random graphs, whose degree distribution is a Poisson distribution of average $z$. In this case we take a population of $\mathcal{N}_{pop}= 10^{3}-10^{5}$ cavity marginals to perform the population dynamics in the single-link approximation. At each update iteration, a degree value $k$ is drawn from the degree distribution of the random graph under study and $k$ messages are chosen randomly from the populations. One the $k$ messages is replaced by the value computed using the remaining $k-1$
messages as input of the BP equations. The update rule is iterated till convergence. Since evaluating the convergence of the whole population of messages is computationally demanding, we assumed a convergence criterion based on a global observable. More precisely, we required that the difference between the computed values of the average activation time $\tau = \sum_t t P(t)$ before and after a sweep of updates over the whole population is smaller than a fixed tolerance value (we fixed this value to be $10^{-3}$). The results, for average degree $z=5$ and threshold values $\theta_i = \lfloor (k_i + 1)/2 \rfloor$, are shown in  Figure \ref{figRGz5}. For $\epsilon = 0.2$, the BP equations converge in the whole range of values of $\mu$ and we obtain a slight optimization of the trajectories compared to the random case ($\epsilon=0$).  Increasing $\epsilon$, convergence issues are possible. However, on given instances of ER random graphs, one can successfully use the MS algorithm, that is able to activate the full system with a density of seeds much lower than the values obtained using the single-link BP approach (see the arrow in Fig.\ref{figRGz5}).

\section{Conclusions}

The study of inverse dynamical problems on large graphs provides a new and exciting application of message-passing algorithms. The use of single-time cavity marginal for the study of progressive dynamics from randomly distributed seeds is fairly general and it can be straightforwardly applied to models with stochasticity. On the contrary the powerful two-times cavity method developed for the inverse dynamical problem is limited to deterministic settings, such as the LTM, that already include a series of relevant real-world problems. The analysis of large deviations in stochastic progressive models, such as the Independent Cascades model or the Susceptible-Infected epidemic model, can be achieved coupling the current representation with the stochastic optimization approach based on a multi-level message-passing construction proposed in \cite{stochmatching}.
The present analysis paves the way for the derivation of efficient message-passing algorithms for the study of dynamical optimization problems. In fact, the zero-temperature limit of the BP equations provides a Max-Sum algorithm that can be used to find optimal configurations of seeds ensuring a desired final state, with applications to the design of cost-efficient viral marketing campaigns in social networks and optimal vaccination strategies against epidemic spreading. On the one hand the fact that a Max-Sum algorithm takes into account all dynamical constraints makes it more powerful than any centrality-based heuristics usually considered, but its distributed nature makes it much faster than centralized optimization methods based on linear programming and simulated annealing.

\begin{acknowledgments}
The authors acknowledge the european grants FET Open 265496, ERC 267915 and Italian FIRB Project RBFR10QUW4.
\end{acknowledgments}

\appendix
\section{BP equations in the cavity-times representation}\label{BPothasasa}
Recent works concerning the zero-temperature dynamics of the random field Ising model \cite{OS10} and  the susceptible-infected model of epidemic spreading \cite{KN10} have suggested that in the absence of optimization, i.e. when the seeds are randomly drawn,  the dynamics can be correctly analyzed using only single-time cavity marginals. In fact, when $\epsilon_i =0, \forall i\in V$, the two-times formalism that we introduced can be reduced to the set of equations \eqref{eq-chi} for $\chi_{i \ell}(t) = Pr_{-\ell}\{x_i^t = 1\}$, i.e. the probability that node $i$ is active at time $t$ in the absence of the neighbor $\ell$.

In order to show the equivalence, it is convenient to introduce a new representation of the dynamic rule. For each directed edge $(i,\ell)$, the variable $t_{i \ell}$ represents the time at which node $i$ would activate in the absence of node $\ell$, and for $i\notin S$, it satisfies the iterative equation
\begin{equation}
t_{i\ell} = \min_{\sum_{j\in \dd i\sm \ell}  w_{ji} \1[t_{j i}<t] \geq \theta_i} t \,,
\end{equation}
whereas $t_{i \ell}=0$ $\forall \ell\in \partial i$ if $i\in S$. As it happens for the activation times $\{t_i\}$, also the equations for the cavity-times admit a unique solution for a given choice of $S$, which is in one-to-one correspondence with the solution of the single $t_i$ model.
For convenience, let us define:
\begin{subequations}
\begin{align}
 f_{i}\left(\left\{ t_{k}\right\}_{k \in \dd i} \right)  & =   \min\left\{ t:\sum_{k\in \dd i}w_{k i}1\left[t_{k}<t\right]\geq\theta_i\right\} \\
 f_{ij}\left(\left\{ t_{k}\right\}_{k \in \dd i\sm j} \right)  & =   \min\left\{ t:\sum_{k\in \dd i \sm j}w_{k i}1\left[t_{k}<t\right]\geq\theta_i\right\}
\end{align}
\end{subequations}
when $i$ is not a seed, $t_{i\ell}$ satisfies the iterative equations $t_{i\ell} =  f_{i\ell}\left(\left\{ t_{k}\right\}_{k\in \dd i\sm \ell} \right)$.
In order to optimize the trajectories and average over the initial conditions,  we introduce the messages $\hat{H}_{i \ell}(t_{i\ell},t_{\ell i})$, defined over the cavity times, that represent the joint probability that $i$ and $\ell$ would activate at times respectively $t_{i\ell}$ and $t_{\ell i}$ in the absence of the other. The messages satisfy the following BP equations
\begin{eqnarray*}
\hat{H}_{i\ell}\left(t_{i\ell},t_{\ell i}\right) & \propto & \sum_{\left\{ t_{ki},t_{ik}\right\} _{k\in \dd i\setminus\ell}}\prod_{k\in\dd i\setminus\ell}\hat{H}_{ki}\left(t_{ki},t_{ik}\right)\left\{ \prod_{k\in \dd i}\delta\left(t_{ik},f_{ik}\left(\left\{ t_{k'i}\right\} _{k'\in i\setminus k}\right)\right)e^{-\epsilon_i\delta\left(f_{i}\left(\left\{ t_{ki}\right\} _{k\in \dd i}\right),\infty\right)}+\prod_{k\in \dd i}\delta\left(t_{ik},0\right)e^{-\mu_i}\right\} \\
 & = & \sum_{\left\{ t_{ki}\right\} _{k\in i\setminus\ell}}\delta\left(t_{i\ell},f_{i\ell}\left(\left\{ t_{ki}\right\}_{k\in \dd i\setminus\ell}\right)\right)e^{-\epsilon_i\delta\left(f_{i}\left(\left\{ t_{ki}\right\}_{k\in \dd i}\right),\infty\right)}\prod_{k\in \dd i\setminus\ell}\hat{H}_{ki}\left(t_{ki},f_{ik}\left(\left\{ t_{k'i}\right\} _{k'\in \dd i\setminus k}\right)\right)+\\
 &  & +\delta\left(t_{i\ell},0\right)\prod_{k\in \dd i\setminus\ell}\hat{H}_{ki}\left(t_{ki},0\right)e^{-\mu_i}.
\end{eqnarray*}
When $\epsilon_i = 0$ $\forall i$, the hypothesis that the messages $\{\hat{H}_{i \ell}(t_{i \ell },t_{\ell i })\}$ do not depend on the backward cavity times $\{t_{\ell i}\}$ is self-consistently satisfied, and the BP equations can be easily reduced to single-time quantities $\hat{H}_{i \ell}(t_{i \ell})$. To show this, we assume on the r.h.s. that $\hat{H}_{k i}\left(t_{ki},y\right)=\hat{H}_{k i}\left(t_{ki}\right)$ and  we get that $\hat{H}_{i\ell}\left(t_{i\ell},t_{\ell i}\right)$ does not depend on the second argument, i.e.
\begin{equation}
\hat{H}_{i\ell}\left(t_{i\ell}\right) \propto\sum_{\left\{ t_{ki}\right\} _{k\in\dd i\setminus\ell}}\left\{ \delta\left(t_{i\ell},f_{i\ell}\left(\left\{ t_{ki}\right\} _{k\in\dd i\setminus\ell}\right)\right)+\delta\left(t_{i\ell},0\right)e^{-\mu_i}\right\} \prod_{k\in\dd i\setminus\ell}\hat{H}_{ki}\left(t_{ki}\right).
\label{eq:HHupdate}
\end{equation}
This implies that the hypothesis is self-consistent and will be verified at every iteration if it is verified at the initial one (e.g. if the messages have uniform initialization).

We need now to consider time-cumulative quantities, such as the probability $\chi_{i\ell}(t)=\sum_{t_{i\ell}\leq t}\hat{H}_{i\ell}(t_{i\ell})$
that node $i$ is active at time $t$ in the absence of node $\ell$. Let us first define the following sequence of increasing sets $U_t$
\begin{align*}
V_{t} & =\left\{ \left\{ t_{ki}\right\} :\sum_{k\in\dd i\setminus\ell}w_{ki}\1\left[t_{ki}<t\right]\geq\theta_{i}\wedge\sum_{k\in\dd i\setminus\ell}w_{ki}\1\left[t_{ki}<t-1\right]<\theta_{i}\right\} \\
U_{t} & =\left\{ \left\{ t_{ki}\right\} :\sum_{k\in\dd i\setminus\ell}w_{ki}\1\left[t_{ki}<t\right]\geq\theta_{i}\right\} \\
U_{t+1} & =U_{t}\cup V_{t+1}\\
U_{t}\cap V_{t+1} & =\emptyset\\
\1\left[U_{t}\right] & =\sum_{0<t'\leq t}\1\left[V_{t'}\right]\\
& = \sum_{0<t'\leq t} \delta\left(t', f_{i\ell}(\left\{t_{ki}\right\}_{k\in \dd i\setminus \ell})\right)
\end{align*}

Let us compute the time-cumulative quantities:
\begin{align*}
\chi_{i\ell}(t) & =\sum_{t_{i\ell}\leq t}\hat{H}_{i\ell}\left(t_{i\ell}\right)\\
 & \propto\sum_{\left\{ t_{ki}\right\} _{k\in\dd i\setminus\ell}}\left\{ \sum_{0<t_{i\ell}\leq t}\delta\left(t_{i\ell},f_{i\ell}\left(\left\{ t_{ki}\right\} _{k\in\dd i\setminus\ell}\right)\right)+e^{-\mu_i}\right\} \prod_{k\in\dd i\setminus\ell}\hat{H}_{ki}\left(t_{ki}\right)\\
  & =e^{-\mu_i}+\sum_{\left\{ t_{ki}\right\} _{k\in\dd i\setminus\ell}} \1\left[\sum_{k\in\dd i\setminus\ell}w_{ki}\1\left[t_{ki}<t\right]\geq\theta_{i}\right] \prod_{k\in\dd i\setminus\ell}\hat{H}_{ki}\left(t_{ki}\right)\\
   & =e^{-\mu_i}+\sum_{\left\{ t_{ki}\right\} _{k\in\dd i\setminus\ell}}\sum_{\left\{ x_{k}=0,1\right\} }\prod_{k\in\partial i\setminus\ell}\delta\left(x_{k},\1\left[t_{ki}<t\right]\right) \1\left[\sum_{k\in\dd i\setminus\ell}w_{ki}\1\left[t_{ki}<t\right]\geq\theta_{i}\right] \prod_{k\in\dd i\setminus\ell}\hat{H}_{ki}\left(t_{ki}\right)\\
    & =e^{-\mu_i}+\sum_{\left\{ x_{k}=0,1\right\} } \1\left[\sum_{k\in\dd i\setminus\ell}w_{ki}x_{k}\geq\theta_{i}\right] \prod_{k\in\dd i\setminus\ell}\sum_{t_{ki}}\hat{H}_{ki}\left(t_{ki}\right)\delta\left(x_{k},\1\left[t_{ki}<t\right]\right)\\
     & =e^{-\mu_i}+\sum_{\left\{ x_{k}=0,1\right\} }\1\left[\sum_{k\in\dd i\setminus\ell}w_{ki}x_{k}\geq\theta_{i}\right] \prod_{k\in\dd i\setminus\ell}\left\{x_{k}\sum_{t_{ki}<t}\hat{H}_{ki}\left(t_{ki}\right)+\left(1-x_{k}\right)\sum_{t_{ki}\geq t}\hat{H}_{ki}\left(t_{ki}\right)\right\}\\
      & =e^{-\mu_i}+\sum_{\left\{ x_{k}=0,1\right\} }\1\left[\sum_{k\in\dd i\setminus\ell}w_{ki}x_{k}\geq\theta_{i}\right] \prod_{k\in\dd i\setminus\ell}\left\{ x_{k}\chi_{ki}\left(t-1\right)+\left(1-x_{k}\chi_{ki}\left(t-1\right)\right)\right\}
      \end{align*}
Denoting by $p_{i}=e^{-\mu_i}/(1+e^{-\mu_i})$ the probability of choosing $i$ as a seed and the initial conditions are $\chi_{i\ell}(0)=p_{i}$ $\forall i$ and fixing the normalization factor $(1+e^{-\mu_i})^{-1}=1-p_i$, we obtain
\begin{equation}
\chi_{i\ell}(t+1)=p_{i}+(1-p_{i})\sum_{\{x_{k}=0,1\}}\1[\sum_{k\in\dd i\sm\ell}x_{k}w_{ki}\geq\theta_{i}]\prod_{k\in\partial i\sm{\ell}}\left[x_{k}\chi_{ki}(t)+(1-x_{k})(1-\chi_{ki}(t))\right]\label{os_eq}
\end{equation}
from which we can compute finally the density of active nodes $\rho(t) = \frac{1}{N}\sum_i \rho_i(t)$ (see also \eqref{eq-rho}). Apart from the obvious differences in the details of the dynamical update, the single-link version of \eqref{os_eq} (assuming all $\chi_{ki}$ identical) is substantially equivalent to Eq. (4) in \cite{OS10}.

We remark that the two-times joint probability introduced in our BP formulation is more than a technical artifact; on the contrary, it is quite crucial to allow information to flow backwards in time when optimizing over the final state ($\epsilon_i > 0$).

\section{Relation to the dynamic cavity equations}
Inspired by previous works combining dynamical mean-field theories (such as the dynamical replica theory and generating functional approaches) with the cavity method \cite{HWP04,HPCS05,MC08}, several authors have recently introduced a general formalism to study non-equilibrium dynamical processes on sparse graphs under the name of {\em dynamic cavity method} \cite{KM11,NB09,AM11}. The dynamical cavity method has also strong mathematical similarities with the application of the cavity analysis to the path-integral representation of quantum spin systems \cite{LSS08, KRSZ08}.

It is easy to show that in the absence of optimization ($\epsilon_i =0, \forall i\in V$), the belief-propagation equations presented in the main text can be viewed as a simplified version, valid only for microscopically irreversible processes, of the dynamic cavity equations. To this end, we adopt a formulation similar to that used by Neri and Boll\'e \cite{NB09} by considering the path probability $P(\underline{{\bx}}^t | \underline{{\bh}}^{t})$ of a  trajectory $\underline{{\bx}}^t = \{\bx^0, \dots, \bx^{t}\}$  with $\bx^t = \{x_1^t, \dots, x_{N}^{t}\}$ in the presence of an external field $\underline{\bh}^t = \{\bh^0, \dots, \bh^{t}\}$. On a cavity graph, in which node $i$ and its interactions are removed, the path probability can be written as
\begin{equation}\label{pathprob}
P(\underline{\bx}^t | \underline{\bh}^t ) = P_{(i)}(\underline{\bx}^t | \underline{\bh}^t + \underline{\bu}_{(i)}^t) \prod_{s=1}^{t} W\left[ x_i^{s} | \bx^{s-1}; h_i^s \right] p_0(x_{i}^0)
\end{equation}
where $\underline{\bu}_{(i)}^t$ is an auxiliary external field acting over the neighbors of $i$ and introduced to keep track of the directed influence of $i$ over them along the dynamics.
For simplicity we have taken a factorized distribution over the initial conditions.
By summing \eqref{pathprob} over all possible trajectories of all nodes $j \neq i$, we get the local marginal $P_{i}(\underline{x}_i^t | \underline{\bh}^t + \underline{\bu}_{(i)}^t)$ that satisfies the equation
\begin{eqnarray}
\nonumber P_{i}(\underline{x}_i^t | \underline{\bh}^t + \underline{\bu}_{(i)}^t) & = & \sum_{\underline{\by}_{\partial i}} P_{(i)}(\underline{\by}_{\partial i}^t | \underline{\bh}^t + \underline{\bu}_{(i)}^t) \prod_{s=1}^{t} W\left[ x_i^{s} | \bx^{s-1}; h_i^s \right] p_0(x_{i}^0) \\
& = & \sum_{\underline{\by}_{\partial i}} P_{(i)}(\underline{\by}_{\partial i}^{t-1} | \underline{\bh}^{t-1} + \underline{\bu}_{(i)}^{t-1}) \prod_{s=1}^{t} W\left[ x_i^{s} | \bx^{s-1}; h_i^s \right] p_0(x_{i}^0)
\end{eqnarray}
where $\by_{\partial i}^t$ is the joint trajectory of the neighbors of $i$ up to time $t$, and the last passage comes from the fact that the dynamics is parallel and the transition probability for variable $i$ at time $t$ depends only on the states of neighbors at time $t-1$.
Exploiting the tree-like assumption, we can express the path probability distribution over the neighbors of $i$ in factorized form
\begin{equation}
P_{(i)}(\underline{\by}_{\partial i}^{t-1} | \underline{\bh}^{t-1} + \underline{\bu}_{(i)}^{t-1}) \propto \prod_{j\in \partial i} P_{j i}(\underline{x}_{j}^{t-1} | \underline{h}_{j}^{t-1} + \underline{u}_{ji}^{t-1})
\end{equation}
In the present case, we have $h_i^s = \theta_i$ and $u_{j i}^s = w_{ij} x_i^{s-1}$ for all times $s>0$, while the transition probability is given by the deterministic update rule as follows
\begin{eqnarray}
W\left[ x_i^{s} = 0 | \bx^{s-1}; h_i^s \right] & = & \mathbbm{1}\left[\sum_{k \in \partial i} w_{k i} x_k^{s-1} < \theta_i\right],\\
W\left[ x_i^{s} = 1 | \bx^{s-1}; h_i^s \right] & = & \mathbbm{1}\left[\sum_{k \in \partial i} w_{k i} x_k^{s-1} \geq \theta_i\right].
\end{eqnarray}
For a given directed edge $(i,j)$, thus the local field can be univocally parametrized in terms of the variable in the removed node, leading to the following set of recursive equations for the cavity marginals $P_{i\ell}(\underline{x}_i^t |  \underline{h}_{i}^{t}+\underline{u}_{i\ell}^{t}) \equiv P_{i\ell}(\underline{x}_i^t,  \underline{x}_{\ell}^{t})$,
\begin{equation}\label{dyncav}
P_{i \ell}(\underline{x}_{i}^t,  \underline{x}_{\ell}^t)  \propto  p_0(x_i^0) \sum_{\{\underline{x}_{j}^{t-1}\}_{j \in \partial i\sm\ell}} \prod_{s=1}^{t} W\left[ x_i^s | \bx^{s-1}; \theta_{i} \right]  \prod_{j\in \partial i\sm \ell} P_{j i}(\underline{x}_j^{t-1}, \underline{x}_i^{t-1})
\end{equation}
A path probability is now a vector of $O(4^{t+1})$ variables, but further reduction is possible because the dynamics is microscopically irreversible. The full sequence of $t+1$ binary values taken by $\underline{x}_i^t = \{x_i^0, \dots, x_i^t\}$ can be encoded in a single integer $t_i = \{0, \dots, t\}$ representing the time at which the variable $i$ flips from 0 to 1 (with the convention $t_i = 0$ for a seed). With this parametrization, the dynamic cavity equations \ref{dyncav}  take the form
\begin{eqnarray}\label{dyncavt}
\nonumber P_{i \ell}(t_{i}, t_{\ell})  & \propto &  p_0(\delta_{t_i,0}) \sum_{\{t_{j}\}_{j \in \partial i\sm\ell}}  \mathbbm{1}\left[\sum_{j \in \partial i} w_{j i} \mathbbm{1}[t_j < t_i-1] < \theta_i\right]   \mathbbm{1}\left[\sum_{j \in \partial i} w_{j i} \mathbbm{1}[t_j < t_i] \geq \theta_i\right] \prod_{j\in \partial i\sm \ell} P_{j  i}(t_j, t_i)\\
& = &  e^{- c_i \delta_{t_i,0}} \sum_{\{t_{j}\}_{j \in \partial i\sm\ell}}  \mathbbm{1}\left[\sum_{j \in \partial i} w_{j i} \mathbbm{1}[t_j < t_i-1] < \theta_i\right]   \mathbbm{1}\left[\sum_{j \in \partial i} w_{j i} \mathbbm{1}[t_j < t_i] \geq \theta_i\right] \prod_{j\in \partial i\sm \ell} P_{j  i}(t_j, t_i)
\end{eqnarray}
where we represented the initial factorized distribution in terms of weights over the seeds, i.e. $p_0(\delta_{t_i,0}) = e^{- \mu_i \delta_{t_i,0}}/(1+e^{-  \mu_i \delta_{t_i,0}})$.
It is easy to check that \eqref{dyncavt} corresponds to the BP equations derived in the main text for the cavity messages $H_{i \ell}(t_i,t_{\ell})$ in the case in which all $\epsilon_i=0$.
From the above derivation it is also evident that optimization could be included by introducing in the dynamic cavity equation an additional local energy term $ - \sum_i \sum_{s \leq t} \epsilon_i \delta_{x_i^t,1}$ in order to account for optimization over the trajectories, that in the activation time representation becomes  $\sum_i \epsilon_i \delta_{t_i,\infty}$.

It could be in principle possible to relax the assumption of complete irreversibility, using probability distributions over time-dependent paths, but the optimization of fully reversible dynamics is currently numerically unfeasible and it requires new separate ideas to overcome current computational limitations.

\end{document}